\documentclass[a4paper,12pt]{article}
\usepackage{amsfonts, amsmath, amssymb, amsthm}
\usepackage[toc,page]{appendix}
\usepackage[blocks]{authblk}
\usepackage{array}
\usepackage[english]{babel}
\usepackage{bm}
\usepackage{booktabs}
\usepackage{chngcntr}
\usepackage{enumerate}
\usepackage{epsfig}
\usepackage{float}
\usepackage{fontenc}
\usepackage{graphicx}
\usepackage[hmargin=1in,vmargin=2.5cm]{geometry}
\usepackage[colorlinks,allcolors = blue]{hyperref}
\usepackage{lscape}
\usepackage{mathrsfs}
\usepackage{multicol}
\usepackage{multirow,bigdelim}
\usepackage{rotating}
\usepackage{subfigure}
\usepackage{verbatim}
\usepackage[svgnames]{xcolor}

\usepackage{mdframed}
\usepackage{graphicx}

\usepackage[round]{natbib}
\usepackage{bbm}

\def\vc#1{{\textcolor{black}{{#1}}}}

\begin{document}

\title{\vspace{-3cm}Extrapolation of extreme covariates in generalized additive regression using extreme value theory}
\author[$*$]{\small Viviana \textsc{Carcaiso}}
\author[$\dagger$]{\small Sebastian \textsc{Engelke}}
\author[$\ddagger$]{\small Juliette \textsc{Legrand}}
\author[$*$]{\small Thomas \textsc{Opitz}}
\affil[$*$]{Unité Biostatistique et Processus Spatiaux, INRAE, France}
\affil[$\dagger$]{Research Center for Statistics, University of Geneva, Switzerland}
\affil[$\ddagger$]{Univ Brest, CNRS UMR 6205, Laboratoire de Mathématiques de Bretagne Atlantique, France}
\date{}
\maketitle

\vspace{-1cm}
\begin{abstract}\small
\vc{Predictions with covariates in the tails of the covariate distribution often lack accuracy and are highly uncertain but may be of critical importance in many applications. New covariates may even lie beyond the  range of training data. The problem can be particularly critical in environmental contexts, for example with climate-change scenarios, where new covariates represent future, more extreme climate conditions. We here propose novel methods to improve generalized additive models (GAMs) with focus on extreme covariates and covariate extrapolation.  
Adopting a random-design setting, we continuously integrate GAMs for the bulk of covariate distributions with models motivated by multivariate extreme value theory for high covariate values. We consider continuous responses but develop also a new approach for binary responses by assuming a continuous latent response variable. Our framework imposes a specific structure for large values of the covariates motivated by extreme value theory: by combining a transformation to a specific marginal scale with an appropriate link function, the continuous response variable depends linearly on the covariates. In an application to occurrences and sizes of large wildfires in Europe for the period 2008--2023, we explore how the new method can improve predictions, especially during extreme conditions, using 
environmental and meteorological covariates}.\\
\vc{{\footnotesize \textsc{key words:} covariate extrapolation; generalized additive models;  multivariate extreme value theory; predictive modeling.}}
\end{abstract}

\section{Introduction}
We consider a random vector $(Y,\bm X)$ composed of a response variable $Y$ and one or several covariates in the random vector $\bm X$, so that we aim to predict $Y$ conditional on $\bm X$. In this random-design setting, the conditional distribution of $Y$ given $\bm X$ is the optimal predictive distribution that uniquely minimizes the expected loss for proper scoring rules \citep{gneiting2007strictly}. We address the problem of ensuring a reliable model when $\bm X$ is extreme, i.e., when covariates are located in the right and left tails of their distribution, where only few observations are available for training the predictive model. Specifically, we seek model structures for appropriate covariate extrapolation with predictions as accurate as possible when new covariates are more extreme than those observed during training. \vc{A typical example are climate-change applications where new covariates, simulated from scenarios of future climate, can contain more extreme conditions. }

Several issues arise in common prediction frameworks. Most tree-based machine learning methods perform constant extrapolation, i.e., predictions remain unchanged when a covariate $x_j$ exceeds the range of its training data, which is unrealistic in many applications. Examples  include standard classification and regression trees \citep{breiman1984classification}, random forests \citep{breiman2001random, biau2012analysis}, and extreme gradient boosting (xgboost) \citep{chen2015xgboost}. By contrast, generalized linear models (GLMs) can extrapolate in a non-constant way beyond observed covariates,  but they are often too rigid to capture complex nonlinearities. The leverage exerted by extreme observations on estimated coefficients depends strongly on the covariate scale; with heavy-tailed covariates, it can be excessively high for covariate outliers. Semi-parametric basis-function approaches, such as generalized additive models (GAMs) with spline bases, are more flexible and provide non-constant extrapolation. However, fitted spline curves are only weakly constrained near the boundaries of the training data, leading to high uncertainty for extreme covariates. By contrast, non-constant extrapolation is not always desirable. If there are saturation effects in the response to covariates,  constraining extrapolation to be constant makes sense. 

To define suitable model structures for the joint distribution of $Y$ and extreme $\bm X$, we draw on  multivariate asymptotic frameworks commonly used for statistical modeling and tail extrapolation of extreme events. Linear relationship arise between extreme covariates and the response on appropriately chosen marginal scales, thus facilitating interpretation, diagnostics and extrapolation. 
We keep a flexible nonlinear predictor structure in the bulk of the covariate distribution (using GAMs) but transition towards linear behavior in the tails. In this framework, we choose how to coherently combine three elements for reliable covariate extrapolation: the marginal transformations of $\bm X$ and $Y$; the response distribution; the link function relating the linear tail to a parameter of the response distribution. Figure~\ref{fig:scenarios} illustrates several stylized scenarios for extreme covariates  (details in Appendix~\ref{app:scenarios}). The quality of tail extrapolation is highly sensitive to the interplay of these three elements. 

\begin{figure}
    \centering
    \includegraphics[width=\linewidth]{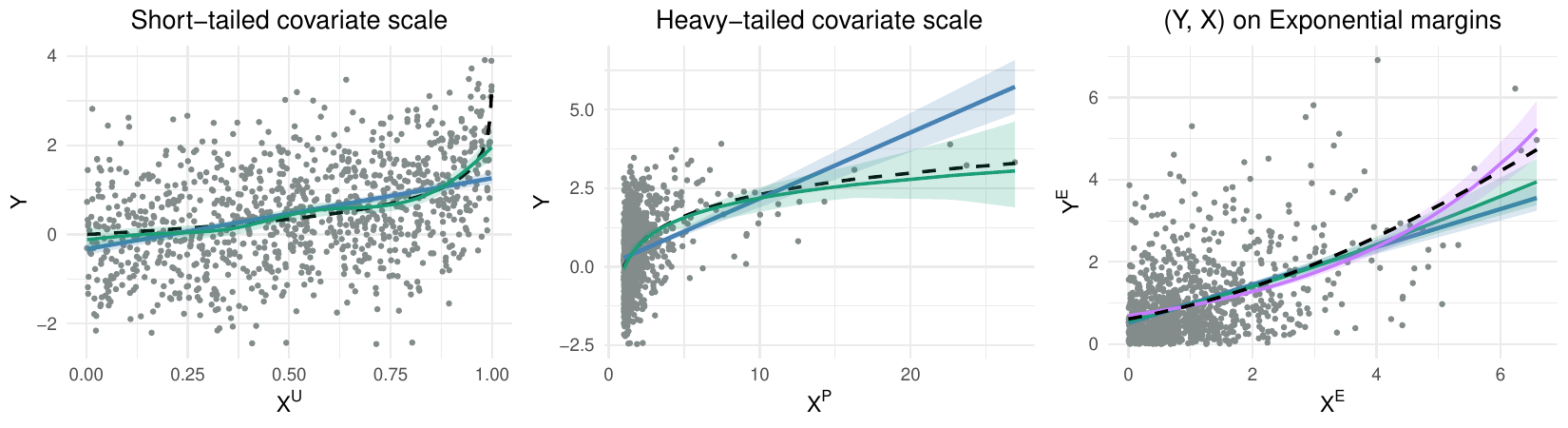}
    \includegraphics[width=\linewidth]{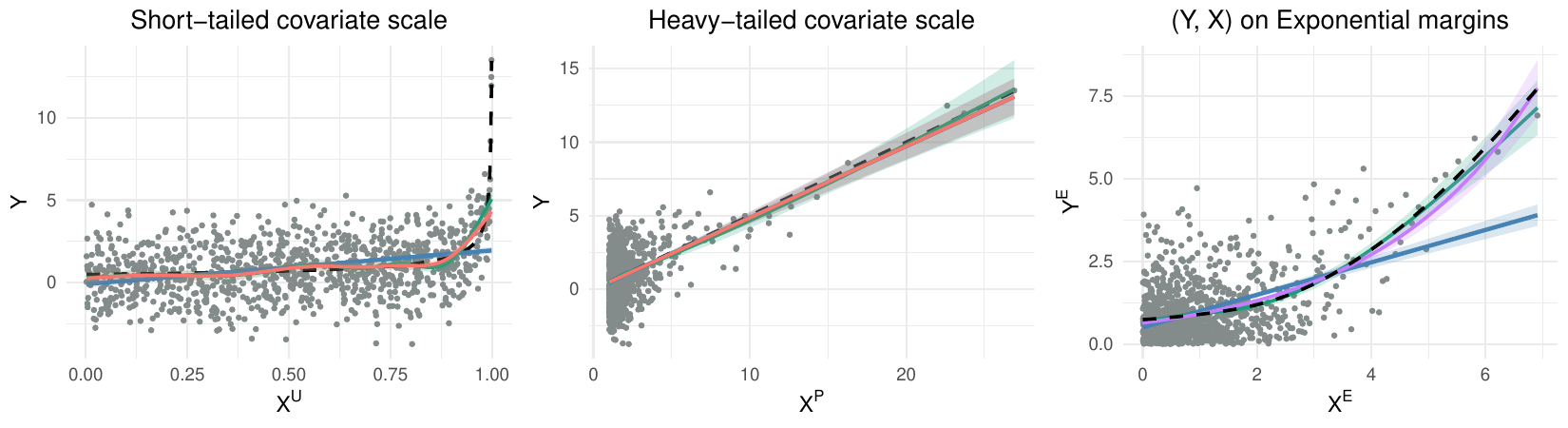}
    \includegraphics[width=\linewidth]{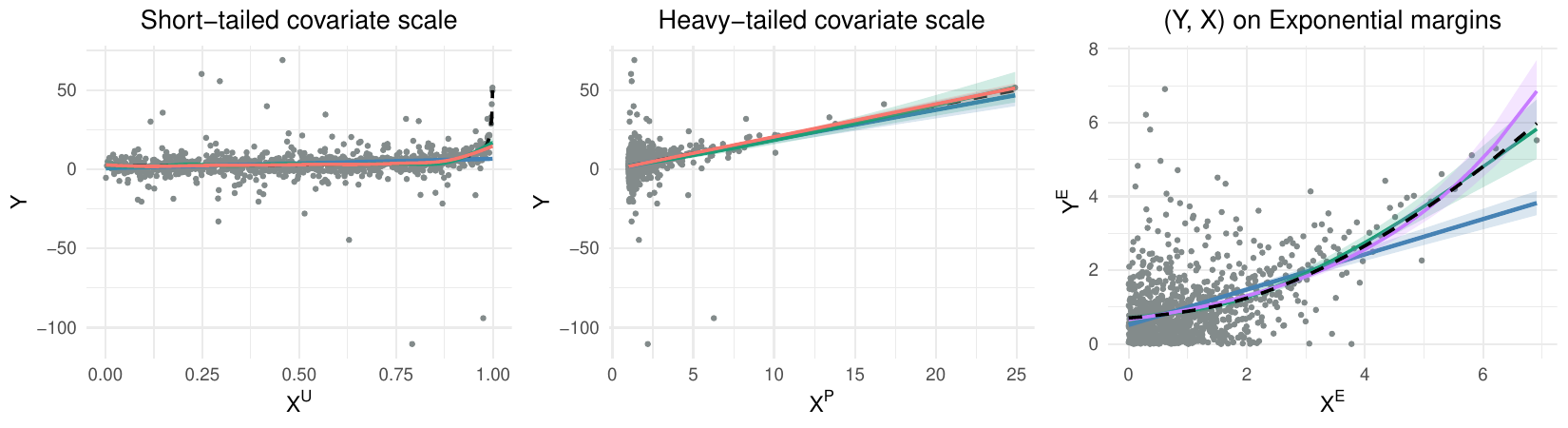}
    \includegraphics[width=\linewidth]{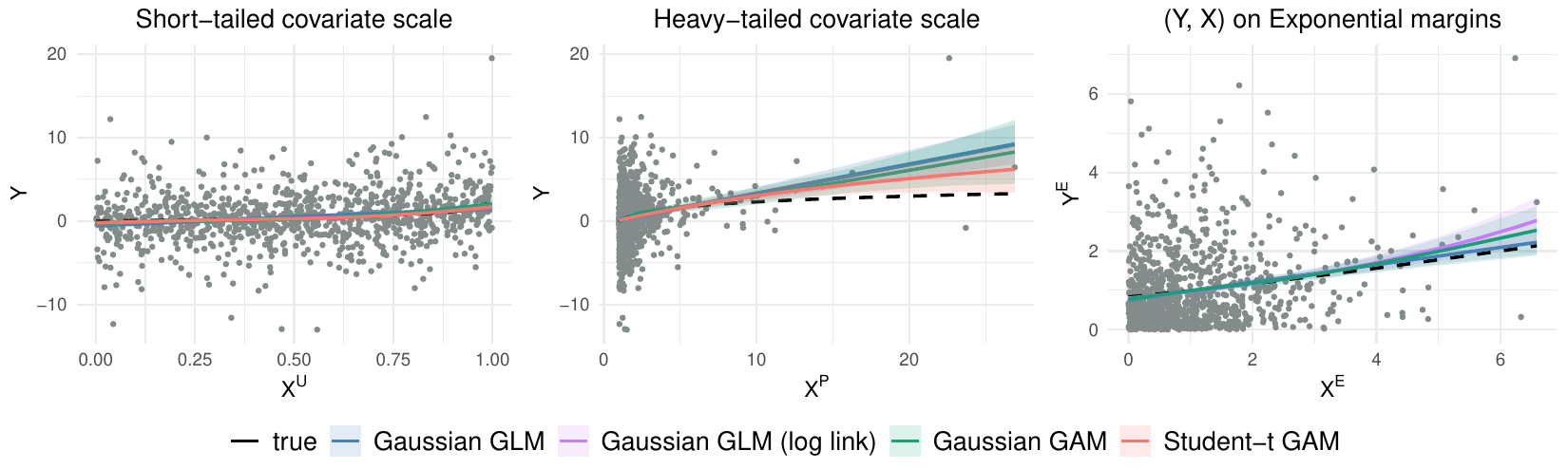}
    \caption{\footnotesize Examples of the effect of covariate and response scaling on model fit under different covariate distributions and error terms. Within each row, the data are identical across columns, but the covariate scale changes (from left to right: uniform on $(0,1)$; Pareto; standard exponential). The third column shows covariate and response on standard exponential margins; see Appendix~\ref{app:scenarios}. Model estimates with 95\% confidence intervals are shown together with the true conditional mean (dashed). The Student-$t$ GAM uses a Student-$t$ error distribution with degrees of freedom $\ge 2$.}
    \label{fig:scenarios}
\end{figure}

We focus on extreme-event frameworks based on threshold exceedances for univariate and multivariate $\bm X$. An extreme multivariate event could be defined as a marginal exceedance above a high threshold ($X_{j_0}>u_{j_0}$), for a fixed component $j_0$ and threshold $u_{j_0}\in\mathbb{R}$, or as an exceedance of a general risk functional ($r(\bm X)>u$), e.g., a norm. Although our theory is developed for the upper tail, it applies similarly to the lower tail, for instance by replacing $\bm X$ with $-\bm X$, or to both tails.

In GAMs, we use a parametric distribution $F_\theta$ with parameter $\theta$ and link function $g$ to obtain the additive structure
\begin{equation}\label{eq:GAM}
 (Y\mid\bm X)\sim F_{\theta(\bm X)}, \quad   g\{\theta(\bm X)\}=\sum_{j=1}^d f_j(X_j),
\end{equation}
with $\bm X$ of dimension $d$. Standard GAMs use linear effects (including an intercept) or smooth basis functions for $f_j$ leading to continuous second-order derivatives of $g\{\theta(\bm X)\}$. 
Covariates may enter several parameters of the response distribution, as in the GAMLSS framework \citep{rigby2005generalized}, where $\theta$ is a vector of parameters with potentially different link functions and linear predictors.

There is growing literature on  modeling  conditional on large values of covariates. The \emph{progression} method of \citet{buritica2024progression} shares similar intuition as our work and focuses on predicting the conditional median with linear covariate extrapolation after suitable marginal transformation,  focusing on random forests and other additive models. They exploit the conditional extremes framework of \citet{heffernan2004conditional} where one conditions on one covariate at a time, and conditioning on several covariates is achieved by iterative backfitting techniques. 
By contrast, we aim to capture the whole conditional distribution of the response, which can be non Gaussian; this requires appropriate distributions for the error terms. The \emph{engression} method of \citet{shen2024engression} also addresses extrapolation with theoretical guarantees for univariate covariates, but considers a framework different from additive noise models such as GAMs.
\citet{clemencon2025weak}  adapt the multivariate regular variation framework of classical multivariate extreme value theory for various machine-learning tasks using the notion of risk; specific uses include \citet{jalalzai2018binary} for binary classification and \citet{huet2023regression} for regression. Multivariate regular variation leads to degenerate representations when dependence between variables is weak; the framework we develop here will be more flexible. We also consider binary responses, a challenging setting since response data carry little information, by using a latent-variable approach rather than classification-based risk \citep{legrand2025evaluation}. Our approach offers a unified view  and review of current methods, and we propose new flexible models to capture both central and extreme covariate behavior.


\subsection{Proposed modeling framework}
\vc{Noting $F_Z$ the distribution of a random variable $Z$, the approach we propose involves standardizing marginal distributions using a quantile function $t^\star$ as follows, 
\begin{align*}
Y^\star = t^\star(F_Y(Y)) ,\quad
X_j^\star = t^\star(F_{X_j}(X_j)),
\end{align*}
where the response variable $Y$ is  assumed to be continuous. We consider the modified GAM
\begin{align}
(Y^\star \mid \bm X^\star) &\sim F^\star_{\theta(\bm X^\star)},\nonumber \\
g\{\theta(\bm X^\star)\} &= \sum_{j=1}^d f_j^\star(X_j^\star),\label{eq:GAM-margins}
\end{align}
where $F^\star_\theta$ is the parametric distribution of $Y^\star \mid \bm X^\star$.
We analyze the structure of models from multivariate extreme value theory for the random vector $(Y^\star, \bm X^\star)$ and use it to derive combinations of $t^\star$, $g$ and $F^\star_\theta$ that allow for coherent predictions at extreme levels of $\bm X^\star$.  We assume that predictor functions $f_j^\star$ are linear in the tail, i.e., above some appropriately chosen threshold $u^\star$, we have 
$$
f_j^\star(x) = f_j(u^\star) + \beta_j (x-u^\star), \quad x>u^\star, 
$$
with slope $\beta_j\in \mathbb{R}$. The resulting predictor  becomes fully linear when $\min_{j=1}^d X_j^\star > u^\star$, i.e., $g\{\theta(\bm X^\star)\} = \tilde{\beta}_0 + \sum_{j=1}^d \beta_j X_j^\star$ with some intercept $\tilde{\beta}_0$.
Some frameworks impose these linearity constraints (on original marginal scale) by default: GLMs are fully linear with $u^\star=-\infty$, and GAMs as implemented in the \emph{R} package \texttt{mgcv} \citep{wood2015package} set $u^\star$ to the maximum training covariate value. Constant extrapolation arising in many machine learning models can be interpreted as $\beta_j=0$. }

\subsection{\vc{Notation and paper organization}}
We adopt the following notation throughout the paper. Vector operations are defined componentwise; for example, $\bm x+\bm y$ has components $x_j+y_j$, and $\bm x>\bm 0$ means that $x_j>0$ for every component $j$. Operations involving a scalar and a vector (or a set) are also understood componentwise (by replicating the scalar). For example, given scalar $t$ and set $B\subset\mathbb{R}^d$, we define $tB=\{tb:b\in B\}$.
Unless stated otherwise, we assume row vectors to avoid cumbersome transpose notation.

The remainder of the paper is organized as follows. Section~\ref{sec:mv-extremes} reviews  frameworks for multivariate extremes and adapts them to construct predictive models with linear structure in the covariate tails. Section~\ref{sec:regression} shows how to combine marginal scale, link function, and response distribution for each framework, while Section~\ref{sec:stat} discusses statistical implementation of the proposed approach. Section~\ref{sec:sim} presents a simulation study. Section~\ref{sec:applications} applies the new methods to wildfire data, and Section~\ref{sec:discussion} concludes with a discussion of the methodology, its limitations, and possible extensions.

\section{Predictive distributions from multivariate extremes frameworks}\label{sec:mv-extremes}

Different frameworks exist to describe the dependence structure of multivariate random vectors, including approaches specific to extreme events. We review, compare, and adapt them to the setting where the vector $(Y,\boldsymbol{X})$ is composed of covariates $\boldsymbol{X}$ and a response $Y$, with extreme events arising specifically in $\boldsymbol{X}$. In this section, $(Y,\boldsymbol{X})$ is assumed to have continuous marginal distributions. Classical multivariate extreme value theory relies on tail regularity assumptions formalized through multivariate regular variation \citep{deHaan2006,Resnick2008}. However, under asymptotic independence when the most extreme possible events in several variables are not simultaneous, the limiting dependence is degenerate, preventing the construction of useful statistical models \citep{Huser2025}. This has motivated the development of more refined approaches that can capture faster joint tail decay rates as often encountered in practice; see the discussion and references in \citet{Huser2025}. 

We focus on four common frameworks for the statistical modeling of extremal dependence. On appropriate marginal scales for $(Y,\boldsymbol{X})$, each framework can induce a linear dependence structure between the response and extreme covariates. Superscripts denote variables transformed to a specific marginal scale; marginal transformations are further discussed in Section~\ref{sec:margins}. For simplicity, we present the theory assuming that all covariates are 
subject to extrapolation.

\subsection{Asymptotic dependence coefficient for the predictive setting}

A common measure for assessing how strongly the extreme events in two variables are associated   is tail correlation \citep{sibuya1960bivariate}, defined for two continuous random variables $Z_1\sim F_1,Z_2\sim F_2$ as the conditional probability limit  $\chi=\lim_{u\rightarrow 1} \Pr(F_1(Z_1)>u\mid F_2(Z_2)>u)\in [0,1]$, if it exists. 
We here propose a variant to evaluate the strength of association in the tail between the response and a predictor integrating covariate information. 

We define the point predictor of a model (where available, as for example in most common GAMs) as the model's best single-value prediction for the response at given covariate values. In GAMs, this is the inverse link function applied to the estimated linear predictor, for example the estimated conditional mean. 
Denote the point predictor $\hat{Y}(\bm X)$ with its distribution function $\hat{F}(y)=\text{Pr}(\hat{Y}(\bm X)\leq y)$, not conditioned on $\bm X$. We introduce the \emph{coefficient of extreme point prediction} defined as the following limit (if it exists) 
\begin{equation}\label{eq:chi-point}
\chi(Y,\hat{Y}(\bm X)) = \lim_{u\rightarrow 1}\text{Pr}(\hat{F}(\hat{Y}(\bm X)) > u \mid F_Y(Y) > u )\in[0,1]
\end{equation}
If $\hat{F}$ is continuous, we can also inverse the conditioning: $\chi(Y,\hat{Y}(\bm X)) = \lim_{u\rightarrow 1}\text{Pr}({F}_Y(Y) > u\mid \hat{F}(\hat{Y}(\bm X)) > u)$. 
This coefficient is in $[0,1]$ and measures the proportion of the most extreme values of the point predictor corresponding to the most extreme events of $Y$. In the case of perfect point predictability where $\hat{Y}(\bm X) \equiv Y$, we have $\chi(Y,\hat{Y}(\bm X))=1$.
If the point predictor provides no information on $Y$, i.e., if it is independent of $Y$, then $\chi(Y,\hat{Y}(\bm X)) = 0$. We describe the strength of asymptotic association  using $\chi(Y,\hat{Y}(\bm X))$, i.e., if $\chi(Y,\hat{Y}(\bm X))>0$ the model exhibits asymptotic dependence, while if $\chi(Y,\hat{Y}(\bm X)) = 0$ we have asymptotic independence. 
The case $\chi(Y,\hat{Y}(\bm X))=0$ does not necessarily imply that the point predictor is uninformative about $Y$; one could further define an analogue of the classical $\bar{\chi}$-measure \citep{coles1999dependence}, refining asymptotic independence by quantifying the rate of joint tail decay between extreme values of $Y$ and the point predictor.

\subsection{Multivariate Gaussian model}\label{sec:mv-gauss}
The multivariate normal distribution is the standard framework for multiple linear regression with random design. While not rooted in extreme value theory, it can serve as a baseline model, with data pretransformed to standard Gaussian margins. It sometimes provides an appropriate fit for relatively weakly dependent multivariate extreme events \citep{Bortot2000}; notice that $\chi(Y^{(G)},\hat{Y}(\bm X^{(G)}))=0$ (except for perfect prediction), i.e., at the most extreme covariate levels, the response will be at a less extreme level due to the additive Gaussian error term.
\vc{Appendix~\ref{app:mv-gaussian} provides technical details.}

\subsection{Conditional extremes model}
The conditional extremes framework \citep{heffernan2004conditional} characterizes the behavior of a vector of variables conditional on another variable exceeding a large threshold. 
We use it to describe the behavior of a response conditioned on a single covariate. Response ${Y}^{(ET)}$ and covariate vector $\boldsymbol{X}^{(ET)}$ are on a common marginal scale ensuring exponential upper tail behavior; more details on exponential-tailed distributions can be found in Appendix \ref{app:marginal}. A marginal probability integral transform is usually required in practice, and it will be discussed later in Section \ref{sec:margins}.
A standard bivariate conditional extremes model is 
\begin{equation}\label{eq:bv-cond-extr}
    \left({Y}^{(ET)}\mid X_{j_0}^{(ET)} > u^{(ET)}\right) 
    = \beta_{j_0} X_{j_0}^{(ET)} + \left(X_{j_0}^{(ET)}\right)^{\gamma} \varepsilon,
\end{equation}
where $\beta_{j_0} \in [-1, 1]$ is a necessary constraint ensuring coherence with the exponential tails. \vc{If $\varepsilon$ has Weibull-type tail, i.e.,  $F_\varepsilon(v) = 1-\exp\{-v^p_\varepsilon\ell(v)\}$ with $\ell$ slowly varying function at infinity, then $\gamma$ must necessarily satisfy $\frac{p_{\varepsilon}}{\gamma(p_{\varepsilon} + 1/\gamma)} \le 1$; e.g., $p_\varepsilon=2$ and $\gamma \in [0,1/2]$ for Gaussian $\varepsilon$, and $p_\varepsilon=1$ and $\gamma=0$ for exponential-tailed $\varepsilon$. See Appendix~\ref{app:gamma-constraint} for the proof of this constraint (to our knowledge not yet known from previous literature).}

In this framework, the tail correlation coefficient between response and point predictor is equivalent to the one between response and single covariate.
The conditional extreme model can describe both asymptotic dependence and asymptotic independence, depending on the values of the parameters. For instance, when $\beta_{j_0}=1$ and $\gamma=0$, then $Y^{(ET)}$ and $X_{j_0}^{(ET)}$ are asymptotically dependent, as well as $Y^{(ET)}$ and its point predictor. For more details on what other values mean we refer to the original paper of \cite{heffernan2004conditional}.

The theory for this framework is directly applicable for conditioning only on a single variable.
With several covariates, one could condition iteratively on each covariate and repeatedly fit the model, always one covariate at a time, as in classical backfitting algorithms \citep{hastie1987generalized}. At each iteration, we could transform the residuals of the previous iteration to the standardized marginal scale before fitting the model, similar to the approach of \cite{buritica2024progression}. 
Instead, we here work directly with multiple covariates, as in modern GAMs \citep{wood2017generalized}.
In the special case where, after applying the conditional extremes model based on one covariate, the response and the remaining covariates are assumed to be jointly normal,
conditioning the response on the additional covariates retains a linear predictor structure and normal residual  distribution, because conditional normal distributions remain normal. To allow using standard GAM implementations (especially Gaussian location-shape families),  we here propose the following variant of the regression model to handle multiple conditioning covariates: 
\begin{equation}\label{eq:CE-multi}
\left({Y}^{(ET)}\mid \bm X^{(ET)} \right)
=  \sum_{j=1}^d\beta_{j} X_{j}^{(ET)} + \left\{ \prod_{j=1}^d\left(X_{j}^{(ET)}\right)^{\gamma_j}\right\} \varepsilon,
\end{equation}
where $\varepsilon \sim \text{N}(0, \sigma^2_{\varepsilon})$\vc{, and $\beta_j\in[-1,1]$ and $\gamma_j\ge 0$ with $\sum_{j=1}^d \gamma_j \leq 1/2$ are necessary parameter constraints that avoid incompatible tails of $Y^{(ET)}$ and the predictive distribution.} 
 
\subsection{Models based on multivariate regular variation}

Multivariate regular variation  is the key concept of classical multivariate extreme value theory \citep[see e.g.,][]{Resnick2008}. 
\vc{To simplify formulas and separate dependence from marginal behavior, one often first applies a marginal transformations ensuring a non negative distribution with standard Pareto tail behavior (see Appendix \ref{app:marginal} for details and examples)}; we write $Y^{(PT)}$ and $X_j^{(PT)}$, $j=1,\ldots,m$, for the standardized variables. 
We use this framework to motivate a regression model for extremes as follows (see  Appendix~\ref{app:mrv} for technical details). We consider a homogeneous risk function $r:[0,\infty)^d\rightarrow [0,\infty)$, continous at the origin, where $r(t\boldsymbol x) = tr(\boldsymbol{x})$ for all $t>0$ and $\boldsymbol x\geq \boldsymbol 0$, and $r(\bm x_n)\rightarrow 0$ if $\min_{j=1}^d x_j\rightarrow 0$; for example, $r$ could be a norm. \vc{Multivariate regular variation ensures the following convergence when marginal distributions are Pareto-tailed:
\begin{equation}\label{eq:Y-given-risk-exceedance}
\left(\frac{Y^{(PT)}}{r(\boldsymbol X^{(PT)})}\mid  r(\boldsymbol X^{(PT)}) > t\right) \rightarrow \varepsilon_r, \quad t\rightarrow \infty, 
\end{equation}
with a nonnegative residual vector $\varepsilon_r$, independent of $\bm X^{(PT)}$, whose distribution $F_{\varepsilon_r}$ depends on $r$.   If $r(\boldsymbol{X}^{(PT)})$ and $Y^{(PT)}$ are asymptotically independent, then $\varepsilon_r\equiv 0$, i.e., the limit is degenerate, which suggests using this framework only when extremal dependence between the predictor $r(\boldsymbol{X}^{(PT)})$ and $Y^{(PT)}$ is rather strong.  
In practice, we can approximate $Y^{(PT)}\approx r(\boldsymbol{X}^{(PT)}) \varepsilon_r$ given that $r(\boldsymbol{X}^{(PT)})$ is relatively large.}




With a single covariate, we can set $r(x)=x$ and obtain $Y^{(PT)}\mid X^{(PT)} \approx X^{(PT)}\times \varepsilon$; 
the framework therefore is quite rigid in this case. 
With several covariates, more interesting linear predictor structures can arise by specific choices of $r$ in \eqref{eq:Y-given-risk-exceedance}. First, consider $r_{\bm \beta}(\boldsymbol x) = \sum_{j=1}^d \beta_j x_j$ with coefficients $\boldsymbol \beta \geq 0$ and $\max_{j=1}^d \beta_j>0$. Then, using \eqref{eq:Y-given-risk-exceedance} and a high threshold $u$ for $r_{\bm \beta}(\bm X^{(PT)})$, we can model 
\begin{equation*}
    (Y^{(PT)} \mid \bm X^{(PT)})    = \left(\sum_{j=1}^d \beta_j X_j^{(PT)}\right) \varepsilon_{\boldsymbol{\beta}}, \quad \text{for } r_{\bm \beta}(\bm X^{(PT)})>u. 
\end{equation*}
The right-hand side of \eqref{eq:Y-given-risk-exceedance} is invariant to rescaling of $\bm \beta$, i.e., when we replace $\bm \beta$ by $t\bm \beta$ for any $t>0$. To compare predictions of $Y^{(PT)}$ given $\bm X^{(PT)}$ across different choices of $\bm \beta$, we could fix the scaling of $\bm \beta$ such that $\lim_{t\rightarrow \infty} t \Pr(r_{\bm\beta}(\bm X^{(PT)}) > t) = 1$. This makes the tail behavior or $r(\bm X^{(PT)})$ invariant to $\bm \beta$, and we can focus on the behavior of $\varepsilon_{\bm \beta}$. To  predict $Y^{(PT)}$ as accurately as possible, we seek the configuration $\bm \beta$ minimizing the residual stochasticity in $\varepsilon_{\bm \beta}$.

Another interesting choice leading to linear covariate representation is the weighted geometric mean  $r(\boldsymbol x) = \prod_{j=1}^d  x_j^{\beta_j}$ with $\boldsymbol \beta \geq 0$ and $\sum_{j=1}^d \beta_j =1$. Then, using log-transformed variables establishes exponentials tails in $Y^{(ET)} = \log Y^{(PT)}$ and $X_j^{(ET)} = \log X_j^{(PT)}$, and we obtain the following linear model  for relatively large $\sum_{j=1}^d \beta_j X_j^{(ET)}$:   
\begin{equation}\label{eq:geom-mean-risk}
   ( Y^{(ET)} \mid \boldsymbol X^{(ET)})    = \sum_{j=1}^d \beta_j X_j^{(ET)} +  \tilde{\varepsilon}_{\boldsymbol{\beta}}.
\end{equation}
\vc{The widely used H{\"u}sler--Reiss model \citep{husler1989maxima,engelke2015estimation} is a special case of the above limits under multivariate regular variation; it is  the extreme value equivalent of multivariate Gaussian distributions. In \eqref{eq:geom-mean-risk}, it leads to a Gaussian distribution with $\tilde{\varepsilon}_{\boldsymbol{\beta}}\sim\mathcal{N}(-\lambda_{\bm \beta}/2,\lambda_{\bm \beta})$ with parameter $\lambda_{\bm \beta}>0$ \citep[see][for technical details]{bolin2025intrinsic}}. The coefficient of extreme point prediction $\chi(Y^{(ET)}, \hat{Y}(\bm X ^{(ET)}))$ is strictly positive since the upper tail of $Y^{(ET)}$ is driven by the  exponential-tailed point predictor $\sum_{j=1}^d \beta_j X_j^{(ET)}$, whereas the Gaussian error is lighter-tailed.
This is in stark contrast to the multivariate Gaussian model, where this coefficient is $0$. When the dependence between response and linear point predictor is relatively weak, the Gaussian or conditional extreme models could be more suitable.

\vc{Finally, some approaches have been developed to define norms and inner products for spaces of multivariate regularly varying distributions, which can then be used to perform best linear prediction based on the projection theorem; see  \citet{cooley2019decompositions,lee2021transformed,cooley_principal_2026,broadhead_vector_2026}, and references therein. This is interesting because linear transformations of independent random variables can approximate arbitrarily closely any asymptotic dependence structure in multivariate regular variation \citep{fougeres2013}. However, theory and implementation are quite technical, and we do not present them here. Taking loose inspiration from this framework, we assume that response ${Y}^{(PT)}$ and covariates $\boldsymbol{X}^{(PT)}$ are univariate Pareto-tailed and have symmetric distribution, and we assume  
\begin{equation*}
    ({Y}^{(PT)} \mid \boldsymbol{X}^{(PT}) = \sum_{j=1}^d \beta_j X_j^{(PT)} + \varepsilon_{\bm\beta},
\end{equation*}
with coefficient vector $\bm \beta=(\beta_1,\ldots,\beta_d)$ and random error variable $\varepsilon_{\bm\beta}$. We call this the regularly-varying linear (\emph{RV-Linear}) model.}

\section{Regression models under different frameworks}\label{sec:regression}

Table~\ref{tab:summary} summarizes the main characteristics of theoretical frameworks from Section~\ref{sec:mv-extremes}, highlighting for each one an appropriate combination of marginal scale, link function and response distribution. Each specific model will be discussed below.

\begin{table}[t]
\resizebox{\textwidth}{!}{
\begin{tabular}{lcccc}
\toprule
 & MV Gaussian & Conditional extremes & H\"usler--Reiss & RV-Linear \\
\midrule
Asymptotic behavior & AI & AI and AD & AD & AD \\
\addlinespace
\multicolumn{5}{l}{\emph{Binary response}}\\
Marginal scale & Normal & Laplace & Laplace & Cauchy\\
Link function & Probit & Logit, Probit & Probit & Cauchit \\
\addlinespace
\multicolumn{5}{l}{\emph{Continuous response}}\\
Marginal scale & Normal & Laplace & Laplace & $t_2$ \\
Conditional distribution & Normal & Normal location--scale & Normal & $t_{\alpha}$ \\
Parameter(s) of interest & Mean & Mean, st. deviation & Mean & Mean \\
Link function & Identity & Identity, log & Identity & Identity \\
\bottomrule
\end{tabular}
}
\caption{\footnotesize Features of different frameworks of multivariate extremes. AD stands for asymptotic dependence and AI for asymptotic independence. $t_{\alpha}$ denotes the Student’s $t$ distribution with $\alpha$ degrees of freedom.}\label{tab:summary}
\end{table}

\subsection{Regression with binary response}\label{sec:binary}\noindent
Let $Y$ represent the binary response variable, where $Y = 1$ indicates an event occurrence, and $\boldsymbol{X}$ be the $d$ dimensional covariate vector. We assume that an event occurs when a continuous latent variable $\Tilde{Y}$ exceeds a threshold $u_Y$:  
\begin{equation*}
Y = 
\begin{cases}
1, & \Tilde{Y} > u_Y, \\
0, & \Tilde{Y} \le u_Y.
\end{cases}
\end{equation*}  
Here, $Y$ can be viewed as a  censored version of $\Tilde{Y}$ with either left- or right-censoring. From our assumption, it follows that
\begin{equation*}
    \theta(\boldsymbol{X}) = \text{Pr}(Y=1 \mid \boldsymbol{X})=\text{Pr}(\tilde{Y}>u_Y\mid \boldsymbol{X}).
\end{equation*}
Therefore, we can apply the frameworks described in Section~\ref{sec:mv-extremes} to the vector $(\Tilde{Y}, \boldsymbol{X})$. Common link functions for binary regression are quantile functions of continuous distributions on $\mathbb{R}$, including the logit function $g\{\theta(\boldsymbol{X})\} = F_{\text{logistic}}^{-1}\{\theta(\boldsymbol{X})\} = \log[\theta(\boldsymbol{X}) /\{1-\theta(\boldsymbol{X})\}]$, the probit function $g\{\theta(\boldsymbol{X})\} = \Phi^{-1}\{\theta(\boldsymbol{X})\}$, and the cauchit function $g\{\theta(\boldsymbol{X})\} = F_{\text{Cauchy}}^{-1}\{\theta(\boldsymbol{X})\}$.
The computations leading to the link functions for each framework results are given in Appendix~\ref{app:link}.

With binary responses, the coefficient  in Eq.~\eqref{eq:chi-point} is supposed to be computed between $\tilde{Y}$ and the point predictor and is a special case of a coefficient proposed by \cite{legrand2025evaluation} for a binary responses and classifiers. 

\subsubsection*{Multivariate Gaussian model}
In the multivariate Gaussian framework (Section~\ref{sec:mv-gauss}), the probability of $Y=1$ can be specified by a GLM with probit link function:
\begin{align*}  
    \Phi^{-1}\left\{\text{Pr}\left(Y=1 \mid \boldsymbol{X}^{(G)}\right)\right\} = - \frac{u_Y}{\sigma_{\varepsilon}} + \sum_{j=1}^{d} \frac{\beta_j}{\sigma_{\varepsilon}}X_j^{(G)} .
\end{align*}

\subsubsection*{Conditional extreme model}
\vc{In the conditional extreme framework, we adopt the common standard Laplace margins (see Appendix~\ref{app:marginal}), ensuring exponential upper tails and symmetry with the lower tail, allowing for handling extremes in both upper and lower tails.}
We can use GLMs to linearly model a function of the conditional probability of success:
\begin{equation*}
    g\{\text{Pr}(Y = 1 \mid X_{j_0}^{(L)} > u^{(L)})\} = \tilde{\beta}_1\left(X_{j_0}^{(L)}\right)^{-\gamma}+\tilde{\beta}_2\left(X_{j_0}^{(L)}\right)^{1-\gamma},
\end{equation*}
where the specific function $g$ depends on the distribution of the residual variable $\varepsilon$. In particular:
\begin{itemize}
  \item $\varepsilon \sim \text{N}(0, \sigma^2_{\varepsilon})$: $g=$ probit, $\tilde{\beta}_1 = -u_Y/\sigma_{\varepsilon}$, $\tilde{\beta}_2 = \beta_{j_0}/\sigma_{\varepsilon}$.
  \item $\varepsilon \sim \text{Laplace}(0, \sigma_{\varepsilon})$: $g=$ logit, $\tilde{\beta}_1 = \text{sgn}(\beta_{j_0})\log2-u_Y/\sigma_{\varepsilon}$, $\tilde{\beta}_2=\beta_{j_0}/\sigma_{\varepsilon}$.
\end{itemize}
For the multivariate version of this model defined in Eq.~\eqref{eq:CE-multi}, GLMs with probit link function can be used to model $g\{\Pr(Y = 1 \mid \bm X^{(L)})\}$, assuming $\gamma_j=0$ for all $j$.

\subsubsection*{Models based on multivariate regular variation}
\vc{We first consider the H{\"u}sler--Reiss model.
For coherence with the conditional extreme framework, we work with variables on Laplace margins. }
Since the conditional distribution of ${Y}^{(L)}$ in this framework is normal and the relationship with the covariates is linear, we can use a GLM with probit link to model the probability of $Y=1$. The mean and variance of $\varepsilon$ are determined by the dependence structure and the coefficient vector $\bm \beta$ (with appropriate constraints), where
\begin{equation*}
     \Phi^{-1}\left\{\text{Pr}\left(Y = 1 \mid \sum_{j=1}^d \beta_j X_j^{(L)} > u^{(L)}\right)\right\} = -\frac{u_Y^{(L)}}{\sqrt{\lambda_{\bm \beta}}} + \frac{\sqrt{\lambda_{\bm \beta}}}{2} + \frac{\sum_{j=1}^d \beta_j X_j^{(L)}}{\sqrt{\lambda_{\bm \beta}}}.
\end{equation*}

Due to the assumption of asymptotic dependence, the H{\"u}sler--Reiss model is restrictive when the dependence is rather weak in the extremes, and the conditional extreme model could be more flexible and covers the H{\"u}sler--Reiss model as a special case by considering $\sum_{j=1}^d \beta_j X_j^{(L)}$ (supposed to be exponential-tailed under this model)  as the conditioning variable $X_{j_0}$ and setting $\beta_{j_0}=1$.

Based on the idea of regulary-varying linear prediction, another option  is the Cauchy distribution, regularly varying with index $\alpha=1$, with cauchit link. 
Since the Cauchy distribution is stable, linear combinations of independent Cauchy variables remain Cauchy distributed. Thus, if $(\tilde{Y},\bm X)=A\bm Z$ are assumed to be related through a linear structure with $A\in\mathbb{R}^{d\times m}$ and an i.i.d. Cauchy vector $\bm Z\in\mathbb{R}^m$, transforming all covariates to Cauchy margins preserves the Cauchy distribution for linear predictors and the error term. 
Therefore, it is interesting to define the Cauchy regression model as
\begin{equation*}
    ({\tilde{Y}}^{(C)} \mid \boldsymbol{X}^{(C)}) = \sum_{j=1}^d \beta_jX_j^{(C)} + \varepsilon, 
    \quad \varepsilon \sim \text{Cauchy}(0, \sigma_{\varepsilon}).
\end{equation*}
Then, 
we can use a binary regression with cauchit link function to model the probability of $Y=1$, with coefficients having the following form:
\begin{equation*}  
F_{\text{Cauchy}}^{-1}\left\{\text{Pr}\left(Y=1 \mid \bm X^{(C)}\right)\right\} =  - \frac{u_Y}{\sigma_{\varepsilon}}+\sum_{j=1}^d \frac{\beta_j}{\sigma_{\varepsilon}}X_j^{(C)} .
\end{equation*}

\subsection{Regression with continuous response}

\subsubsection*{Multivariate Gaussian model}
As mentioned in Section~\ref{sec:mv-extremes}, the conditional model with $(Y^{(G)}, \bm X^{(G)}) \sim \text{N}_{d+1}(\bm 0, \bm \Sigma)$ corresponds to the classical linear model with random covariates. 
What we linearly model is the conditional expectation (or equivalently the median) using the identity link function. This model holds whether or not we are conditioning on the covariates being large, but we can suppose linearity only for extreme $\bm X^{(G)}$.

\subsubsection*{Conditional extreme model}
\vc{For the conditional extreme model, we adopt standard Laplace margins, as done in the binary case. Under this framework,} $\bm \theta( X_{j_0}^{(L)}) = (\mathbb{E}[Y^{(L)}\mid X_{j_0}^{(L)}], \text{sd}[Y^{(L)}\mid X_{j_0}^{(L)}])$, with
\begin{equation*}
    \mathbb{E}[Y^{(L)}\mid X_{j_0}^{(L)}> u^{(L)}] = 
        \beta_{j_0} X_{j_0}^{(L)},
\end{equation*}
and
\begin{equation*}
    \log(\text{sd}[Y^{(L)}\mid X_{j_0}^{(L)}> u^{(L)}]) = 
        \log \sigma_{\varepsilon} + \gamma_{j_0} \log X_{j_0}^{(L)},
\end{equation*}
which is a location-scale model. An identity link is therefore used to model the conditional mean linearly in $X_{j_0}^{(L)}$, and a logarithmic link to model the conditional standard deviation linearly in $\log X_{j_0}^{(L)}$. If $\varepsilon\sim \mathrm{N}(0,\sigma_{\varepsilon})$, then $Y^{(L)}\mid X_{j_0}^{(L)}>u^{(L)}$ belongs to a normal location-scale family.

Analogously, in the multivariate version the conditional expectation is $\sum_{j=1}^d \beta_j X_j^{(L)}$, and the logarithm of the conditional standard deviation is $
        \log \sigma_{\varepsilon} + \sum_{j=1}^d \gamma_j \log X_j^{(L)}$.

\subsubsection*{Models based on multivariate regular variation}
\vc{In the H{\"u}sler--Reiss model, again we assume response and covariates on standard Laplace margins and we model the expectation with identity link function. With a single covariate, we can write the regression model as $g\{\theta( X_{j_0}^{(L)})\} = \theta( X_{j_0}^{(L)}) = \mathbb{E}[Y^{(L)}\mid X_{j_0}^{(L)} > u^{(L)}] = X_{j_0}^{(L)} -\lambda/2$, which is a linear model with regression coefficient fixed to 1 and intercept $-\lambda/2, \lambda>0$. Similarly, in the presence of multiple covariates, the model takes the form}
\begin{equation*}
    g\{\theta(\bm X^{(L)})\} = \theta(\bm X^{(L)}) = \mathbb{E}\left[Y^{(L)}\mid \sum_{j=1}^d \beta_j X_j^{(L)} > u^{(L)}\right] = \sum_{j=1}^{d} \beta_j X_j^{(L)} - \frac{\lambda_{\bm \beta}}{2}.
\end{equation*}

\vc{In the framework of regularly-varying linear prediction where a linear structure arises with heavy-tailed margins, we propose to work with Student’s $t$ margins 
with $\alpha=2$ degrees of freedom, denoted by $t_2$, by analogy with \citet{cooley2019decompositions,lee2021transformed}. We assume an independent error variable $\varepsilon\sim t_\alpha$, $\alpha\geq 2$ (i.e., with similar or relatively lighter tails), a model that is readily available in standard regression software, including the \texttt{mgcv} package in R.
Given $Y^{(t_2)}$ and $\bm X^{(t_2)}$, respectively the response variable and the covariate vector on $t_2$ margins, we assume}
\begin{equation*}
    \frac {Y^{(t_2)} -  \mathbb{E}[Y^{(t_2)}\mid \bm X^{(t_2)}]}{\sigma} \sim t_{\alpha}, \quad \alpha \ge 2,
\end{equation*}
and we model the conditional expectation with a linear predictor using the identity link function, i.e.,
\begin{equation*}
    g\{\theta(\bm X^{(t_2)})\} = \theta(\bm X^{(t_2)}) = \mathbb{E}\left[Y^{(t_2)}\mid \bm X^{(t_2)}\right] = \sum_{j=1}^{d} \beta_j X_j^{(t_2)}.
\end{equation*}

\section{Statistical implementation}\label{sec:stat}

\subsection{Marginal transformations}\label{sec:margins}
Marginal transformations to some standardized scale are useful for theoretical representations and practical interpretations of extreme event distributions as highlighted in Section~\ref{sec:mv-extremes}. Variables can be transformed to a specific marginal scale through a probability integral transform; e.g., if we consider a generic variable $Z\sim F_Z$ with continuous distribution function $F_Z$, then transformation to standard Gaussian margins is achieved by $Z^{(G)}= \Phi^{-1}\{F_Z(Z)\}$, with $\Phi$ the standard normal distribution function. 
To estimate $F_Z$ from a sample $Z_1, \dots, Z_n$, we adopt a common semiparametric modeling approach with nonparametric bulk and parametric tail representation, here applied to both upper and lower tails, defined as
\begin{equation*}
    \hat{F}_Z(z) = \begin{cases}
    \left\{\frac{1}{n}\sum_{i=1}^n\mathbbm{1}\{Z_i> l_Z\}\right\}\left\{ 1+\hat{\xi_l}\frac{l_Z-z}{\hat{\sigma}_l}\right\}_+^{-1/\hat{\xi}_l} & z \le l_Z,\\
        \frac{1}{n}\sum_{i=1}^n\mathbbm{1}\{Z_i\le z\} &  z \in (l_z, u_z],\\
        1- \left\{\frac{1}{n}\sum_{i=1}^n\mathbbm{1}\{Z_i> u_Z\}\right\}\left\{ 1+\hat{\xi}_u\frac{z-u_Z}{\hat{\sigma}_u}\right\}_+^{-1/\hat{\xi}_u} & z> u_Z,
    \end{cases}
\end{equation*}
where $\mathbbm{1}\{\cdot\}$ is the indicator function, $l_Z$ and $u_Z$ are respectively a lower and an upper threshold usually set at a low and a high empirical quantile of $Z$, and $(\hat{\sigma}_l, \hat{\xi}_l)$ and $(\hat{\sigma}_u, \hat{\xi}_u)$ are the estimated parameters of generalized Pareto distributions (GPDs) fitted separately to the lower and upper tail. When the interest is mainly on high values of $Z$, the GPD model is only used for exceedances of $u_Z$, and the empirical distribution function is also adopted in the lower tail. Alternative methods to estimate $\hat{F}_Z$ exist, for instance one could use the extended GPD \citep{naveau2016modeling}.

We transform the response $Y$ and the continuous covariate vector $ \bm X$ to a common marginal scale using the scheme described above. For $\bm X$, this transformation is applied componentwise to each $X_j$, $j=1, \dots, d$. 
Here, we estimate $\hat{F}_Y$ and $\hat{F}_{X_j}$ on all the data, and not separately for training and validation data. The distributions of these two sets could be very different if we (artificially) put more extreme covariate events into the validation set to check extrapolation behavior, as done in the simulation study of Section~\ref{sec:sim}.

\subsection{Linear extrapolation in Generalized Additive Models}\label{sec:splines}
GAMs, as defined in Eq.~\eqref{eq:GAM}, extend a standard linear model by allowing for non-linear functions of each
of the covariates, preserving the additive structure. 
A basis expansion $f_j(X_j) = \sum_{c=1}^{C_j} \beta_{j,c}b_{j,c}(X_j)$ is often used to capture these non-linear relationships between covariates and response, with a very common choice being spline basis functions. 
Our proposed approach entails using GAMs that combine spline function with linear models that arise from the approaches described in Section~\ref{sec:mv-extremes}.
Thus, the spline must be exactly linear after a particular knot/threshold and continuously glued to the flexible part below the threshold.
Linearity after a threshold $u$ means that 
the second derivative vanishes above $u$, and $f_j$ and its first derivative are typically required to be continuous at $u$. 
The strategy we use to enforce linearity after $u$ is to modify the basis and penalties so the tail can only be linear. 
Natural cubic spline bases are constructed to extrapolate  $f_j^\star=s_j$ with a spline function $s_j$ linearly beyond the boundary knots and have smooth second derivates everywhere in $s_j$ \citep{hastie2009elements}. Therefore, we can use $f_j^\star=s_j$ in the GAM in Eq.~\eqref{eq:GAM-margins}  by setting the boundary knot to the threshold $u^\star$. By construction, some covariate values $X_j^\star$ lie beyond the boundary knots, which is by default not admissible in implementations such as the \emph{R} package \texttt{mgcv} but can be bypassed with a custom setup. 
In practice, this approach is numerically stable and straightforward to fit via penalized least squares.


Let $\bm X^\star =( t^\star(F_{X_1}(X_1)), \dots, t^\star(F_{X_d}(X_d)))$, where $t^\star$ is the quantile function of an appropriate marginal scale.
Assuming that we want to model a scalar $\theta(\bm X^\star)$ that can be the conditional expectation or median of the response, 
we can explicitly write the predictor function as
\begin{equation}\label{eq:spline-linear-cases}
    f_j^\star(X^\star_j) = \begin{cases}
        s_j(X_j^\star) &  X_j^\star \le u^\star,\\
        s_j(u^\star) + \beta_{j} (X_j^\star-u^\star), & X_j^\star > u^\star.
    \end{cases}
\end{equation}

If we consider  a penalized cubic regression spline $s_j$, the roughness functional $\mathcal{J}(f_j^\star)$ is defined only on the region below the threshold $u^\star$:
\begin{equation*}
    \mathcal{J}(f_j^\star) = \int_{-\infty}^{u} \{f_j^{\star \prime \prime}(x)\}^2 \, \text{d}x,
\end{equation*}
with continuity constraints at $u^\star$, so that curvature is penalized below the threshold while linearity is enforced above it. Other choices of $s_j$ are possible, such as thin plate splines \citep{duchon2006splines}; for a review of spline functions in R, we refer to \cite{perperoglou2019review}.

The linear tail makes sense when the model for the tail of $X^\star_j$ is exactly linear in $X^\star_j$, which is not true in all frameworks, as illustrated in the Section~\ref{sec:binary}. Moreover, the generalization to multivariate $\bm \theta(\bm X^\star)$, e.g., a location-scale model, is often not straightforward to implement (e.g., in \texttt{mgcv}). Still, it can be used in cases where it is reasonable to assume a conditional extreme model with $\gamma=0$. 

A different approach for achieving the structure in \eqref{eq:spline-linear-cases} can be used if the implementation allows applying the roughness penalty only until $u^\star$ but not beyond. The spline function $s_j(\tilde{X}_j^\star)$ (or any other type of predictor function) is estimated for the modified covariate $X_j^{\star,-}=\min(X_j^\star,u^\star)$, and another covariate $X_j^{\star,+} = \max(X_j^\star-u^\star,0)$ contributes with a linear effect to estimate $\beta_j$. This alternative approach ensures continuity of $f_j^\star$ at $u^\star$, but imposing continuous  derivatives may be more difficult with standard software.
However, it has the benefit that  it can be easily extended to accommodate multivariate $\bm \theta(\bm X^\star)$ and conditional extreme models with $\gamma \neq 0$.

\subsection{Assessment of prediction performance}\label{sec:performance}
To assess model performance, especially for extreme covariates, we use several predictive measures, including the logarithmic score \citep{good1952rational} and the continuous ranked probability score (CRPS) \citep{matheson1976scoring}, two widely used proper scoring rules. More details on proper scoring rules are given in Appendix~\ref{app:scores}.
Weighted proper scoring rules, such as the weighted CRPS of \citet{gneiting2011comparing}, typically emphasize a specific region of the response distribution, for example observations above a high threshold. Since our interest lies instead in the region of high values of the covariates, we define
\begin{equation*}
    S_w(F,y,\boldsymbol{x})=\mathbbm{1}\{r(\boldsymbol{x})>m\}S(F,y),
\end{equation*}
where $r$ is a risk function as previously, $\boldsymbol{x}$ is a realization of the covariate vector, and $m$ is a high threshold chosen to produce enough exceedances to guarantee that the computation of $ S_w(F,y, \boldsymbol{x})$ is still reliable. Weight functions that allow a smooth transition between bulk and right tail of $r(\boldsymbol{x})$ could also be used.

For binary responses, we assess performance using the area under the ROC curve (AUC) \citep[for more details see, for instance,][]{fawcett2006introduction} and the area under the precision--recall curve (AUPRC) \citep{saito2015precision}, which is more appropriate for rare events. Both metrics aggregate model performance across varying classification thresholds, as explained in Appendix~\ref{app:scores}. Their tail-weighted counterparts are defined as
\begin{equation*}
    A_w(\hat p,y,\boldsymbol{x})=\mathbbm{1}\{r(\boldsymbol{x})>m\}A(\hat p,y),
\end{equation*}
where $A$ denotes either AUC or AUPRC, and $\hat p$ is the predicted success probability.

While our modeling focus is on improvements based on weighted measures, gains for extreme covariates should not come at the cost of overall performance. As a complement to the unweighted versions of proper scoring rules, AUC and AUPRC, the Akaike information criterion (AIC) is also included to measure the overall predictive quality of different models.

\subsection{Complete modeling workflow}\label{sec:alg}
\vc{The implementation of the proposed approach to model the distribution
$F^\star_{\theta(\bm X^\star)}$ of $Y^\star \mid \bm X^\star$, with special
attention to extreme covariates, involves the following steps (recall that $\star$ denotes quantities on the transformed/target scale):}

\begin{enumerate}
    \item \vc{\textbf{Marginal transformation.} Estimate $F_{X_j}$, and also $F_Y$
    if $Y$ is continuous, using the semiparametric approach of
    Section~\ref{sec:margins}, and transform to the scale required by the
    chosen framework:
    $X_j^\star = t^\star(\hat{F}_{X_j}(X_j))$ and
    $Y^\star = t^\star(\hat{F}_Y(Y))$, with $t^\star$ the corresponding
    quantile function.}

    \item \vc{\textbf{Train/test split.} Divide the transformed data into
    training and test sets.}

    \item \vc{\textbf{Threshold selection.} Choose a threshold $u^\star$ on the
    transformed scale. This may be a fixed high quantile (e.g., the 95th
    percentile) or a value chosen so that a specific number of observations
    exceeds it (e.g., 500). 
    Preferably, $u^\star$ is chosen by repeating
    Steps 4--5 over a grid of candidate thresholds and selecting the value
    that minimizes a chosen score or the mean rank across several scores.}

    \item \vc{\textbf{Model fitting.} Fit the GAM in \eqref{eq:GAM-margins} to
    the training set with the link $g$ recommended for the chosen
    framework, imposing linearity for extreme covariates:}
    \begin{enumerate}
        \item\vc{Define  $X_j^{\star,-} = \min(X_j^\star, u^\star)$ and
        $X_j^{\star,+} = \max(X_j^\star - u^\star, 0)$, for $j = 1, \dots, d$.
        \item Estimate the spline and linear regression coefficients of
        $$g\{\theta(\bm X^\star)\} = \sum_{j=1}^d s_j(X_j^{\star,-}) + \beta_j X_j^{\star,+},$$
        using the preferred fitting method (e.g., via the package
        \texttt{mgcv}).}
    \end{enumerate}
    \noindent \vc{ \textbf{Alternative implementation of Step 4.} As described in Section \ref{sec:splines}, a smooth transition at the threshold is possible but not available for all model/implementation frameworks. For each $j$:}
\begin{itemize}
    \item[4b$'$)] \vc{Construct a basis for a cubic regression spline $s_j$ by using only the training observations with $X_j^\star < u^\star$ and fixing $u^\star$ as boundary knot. 
    \item[4c$'$)] Fit the model $g\{\theta(\bm X^\star)\} = \sum_{j=1}^d s_j(X_j^\star) $, treating $s_j$ as a penalized term (via its associated smoothing penalty) and adding an unpenalized global linear term in $X_j^\star$ to correct or reinforce the linear behavior in the tail.}
\end{itemize}

    \item \vc{\textbf{Prediction.} Predict the quantity of interest 
    (e.g., the
    probability of success in the binary setting or the average outcome in the continuous setting) for observations in
    the test set. If the response is continuous, transform the predicted values back to the original scale of $Y$. The back-transformation reverses the marginal transformation of Step~1 in two stages: predicted values are first mapped from the transformed scale to the uniform $[0,1]$ scale via $t^{\star-1}$, and then from the uniform scale to the original scale of $Y$ via the inverse of the semiparametric marginal model $\hat{F}_Y$ fitted in Step~1, using empirical quantiles for the bulk, and inverting the GPD for the tail(s).
    \item \textbf{Predictive performance.}  Evaluate predictive performance by
    computing, on the test set, some or all of the measures
    described in Section~\ref{sec:performance}: for a continuous response, scoring rules based on the predicted $\hat{Y}$, and, for a binary response, scoring rules, AUC and AUPRC based on $\hat{\text{Pr}}(Y=1\mid \bm X^\star)$. Threshold-weighted versions of these measures can additionally be computed by restricting the assessment to the $m$ test observations with the largest average rank of $\bm X^\star$. Model comparison in terms of overall fit (e.g., via AIC) can instead be carried out on the training set.
    These measures can then be used to assess and compare candidate models and select the preferred specification.
    \item \textbf{Predictions for new covariates.} To predict $\hat\theta(\bm x^\star)$ for a new covariate vector $\bm x$ (as opposed to observations already available in the test set), apply the same pipeline used above: }
    \begin{enumerate}
        \item \vc{Transform each covariate to the modeling scale using the marginal models estimated in Step~1.
        \item Obtain $\hat\theta(\bm x^\star)$ from the GAM fitted in Step~4. If $\bm x^\star$ falls outside the range of the training covariates, predictions rely on the linear behavior imposed above $u^\star$.
        \item If $Y$ is continuous, back-transform $\hat\theta(\bm x^\star)$ to the original scale of $Y$ as in Step~5.}
    \end{enumerate}
\end{enumerate}

\section{Simulation study}\label{sec:sim}

\subsection{Simulations for binary response}
For simplicity, we focus on a setting with a single covariate. 
To cover a variety of situations, we generate data under three scenarios, selected at random with equal probability, and repeat the simulation to obtain 100 datasets. For each dataset, we draw 10\,000 values of $X$ from a specified distribution and generate $\tilde{Y}=\zeta X  +h(X) + a\varepsilon$, i.e., $\tilde{Y}$ is a combination of a linear term in $X$, a nonlinear component $h(X)$ (here modeled using a spline function), and a stochastic error term $\varepsilon$. The scalar coefficient $\zeta$ is drawn from a uniform distribution, while $a$ is a constant depending on the scenario. Knot locations for the natural cubic spline are obtained by drawing uniformly distributed quantiles and transforming them using the quantile function of $X$, and spline coefficients are drawn independently from $\text{N}(0, 0.49)$. The three scenarios differ in the joint specification of the covariate and error distributions: (i) exponentially distributed covariate with normal noise ($a=1$), (ii) exponentially distributed covariate with Laplace noise ($a=1$), and (iii) a heavy-tailed (Pareto) covariate with heavy-tailed noise (Student's $t$, $a=3$). 

We treat $\tilde{Y}$ as a latent continuous variable, obtaining the binary response $Y$ based on $\tilde{Y}$ exceeding a threshold $u_Y$, such that the observed response conveys only partial information about the latent variable $\tilde{Y}$. The true conditional probability of success is 
\begin{align*}
    \text{Pr}(Y=1 \mid X)&=\text{Pr}(\tilde{Y}>u_Y\mid X)=\text{Pr}(\zeta X + h(X) + a\varepsilon>u_Y\mid X) \\ &= \text{Pr}(\varepsilon>(u_Y - \zeta X - h(X))/a \mid X) = F_{\varepsilon}\left ((\zeta X  + h(X)-u_Y)/a\right ),
\end{align*}
where $F_{\varepsilon}$ is the cdf of $\varepsilon$, 
whose distribution is assumed to be symmetric.
In the specific case of $\varepsilon \sim \text{N}(0, \sigma_{\varepsilon}^2)$, the true model for this probability is a probit model:
\begin{equation*}
    \Phi^{-1}\{\text{Pr}(Y=1 \mid X)\} = \frac{\zeta X}{a\sigma_{\varepsilon}} +\frac{h(X)}{a\sigma_{\varepsilon}} - \frac{u_Y}{a\sigma_{\varepsilon}}.
\end{equation*}

\begin{figure}[b!]
    \centering
    \includegraphics[width=\linewidth]{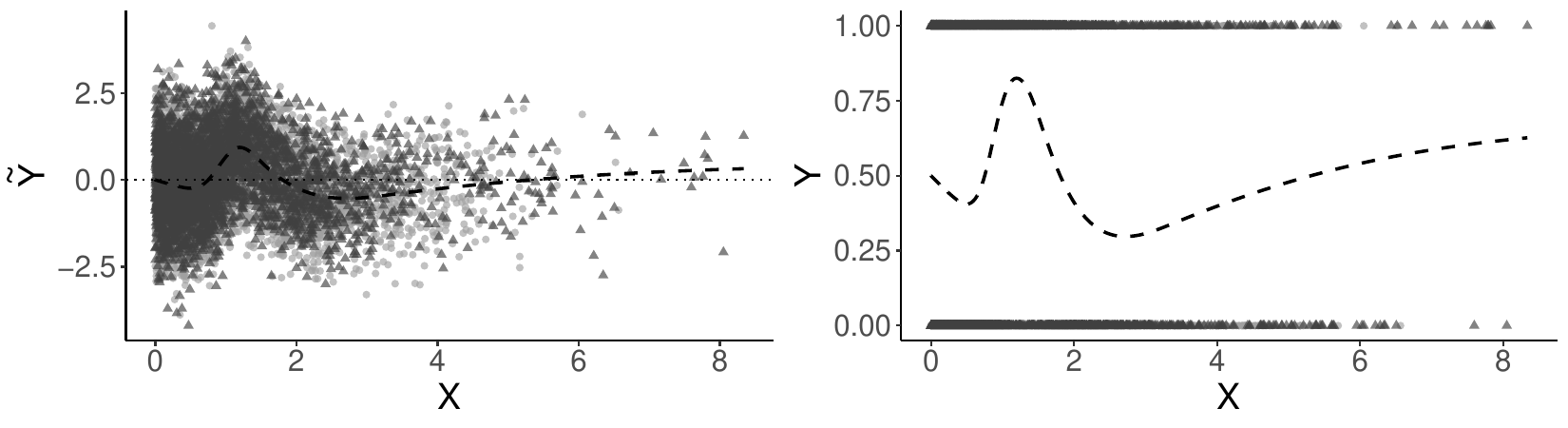}
    \includegraphics[width=\linewidth]{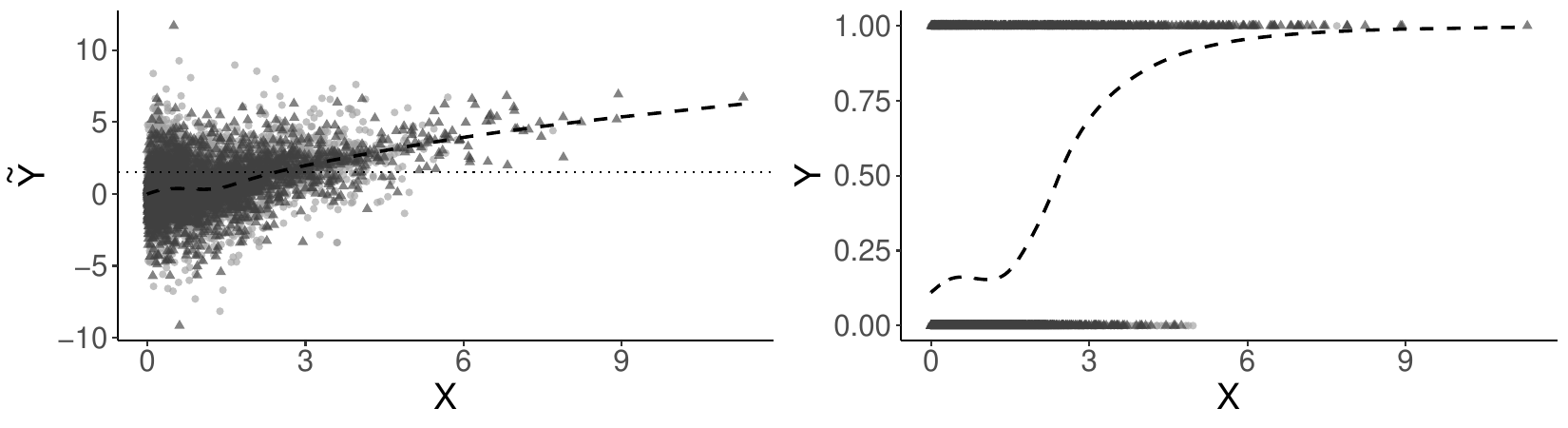}
    \includegraphics[width=\linewidth]{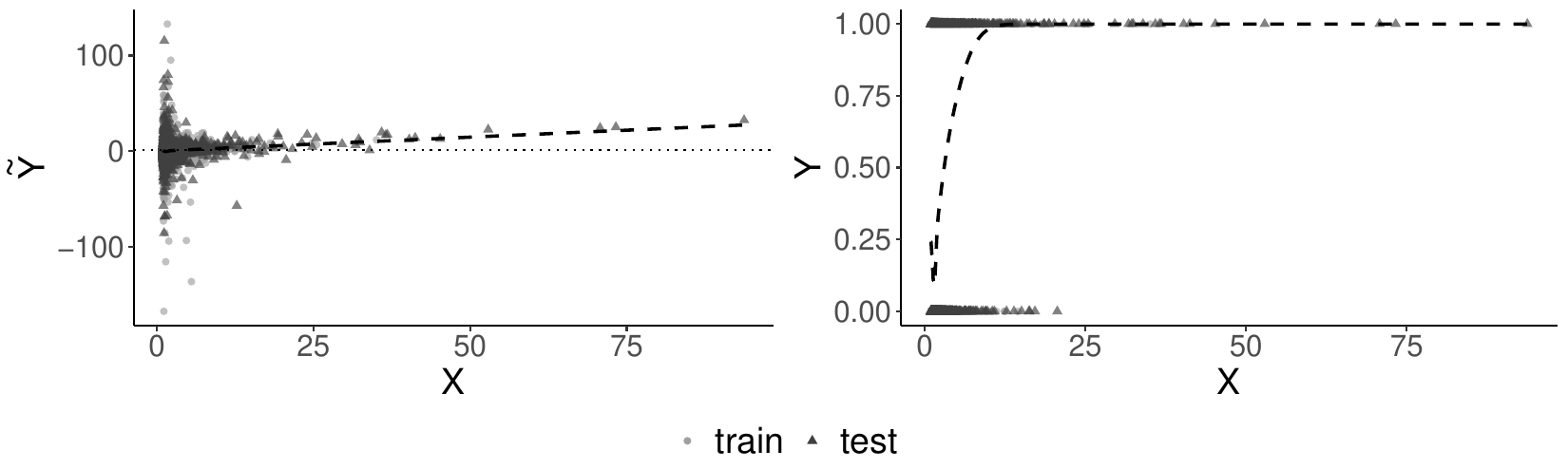}
    \caption{\footnotesize Examples of simulated data under different scenarios. Left: simulated covariate versus latent variable, with the true conditional mean (dashed) and the threshold defining the binary response (dotted). Right: simulated covariate versus binary response, with the true $\Pr(Y=1\mid X)$ (dashed). Different shades and symbols distinguish training and test observations.}
    \label{fig:sim-bin}
\end{figure}

Figure~\ref{fig:sim-bin} shows one simulated dataset for each scenario.
As it can be seen in Figure~\ref{fig:sim-bin}, the observations are divided into training set and test set in a way to have relatively more extreme values of the covariate in the test set. Indeed, the probability that an observation belongs to the test set is defined as
\begin{equation*}
    p_{TEST} = \begin{cases}
        p_0 + \frac{m - (n - r)}{m}  (1-p_0) & \text{if } r\le n-m,\\
        p_0 & \text{otherwise},
    \end{cases}
\end{equation*}
where $p_0$ is the baseline probability of test set (set to 1/3), $m$ is the number of observations considered extreme (in this case the 500 most extreme ones), and $r$ is the rank of the covariate $X$ in the complete covariate sample.

The widely used \emph{R} package \texttt{mgcv} for GAMs imposes  linearity constraints beyond the range of observations by default. Here, we instead control both the tail slope and the threshold at which linearity begins, as defined in Eq.~\eqref{eq:spline-linear-cases}. We compare models with and without this tail correction across different frameworks, using different marginal scales and link functions as explained in Section \ref{sec:regression}.
Instead of fitting the conditional extreme model on the Laplace marginal scale, we use the modified Laplace scale defined in Appendix~\ref{sec:laplace-alt}, referred to simply as Laplace margins from here onward. For these models, we adopt the probit link, thus assuming normal error terms even when the true errors are Laplace or Student-$t$ distributed. This choice ensures coherence and yields a conditional extreme model that is well defined for all selected values of $\gamma$. The construction of the fitted models is agnostic to the structure of simulated data; in particular, we do not exploit any knowledge about spline knots and coefficients, or the choice of residual distribution, used for simulation. 

\begin{figure}[]
    \centering
    \includegraphics[width=\linewidth]{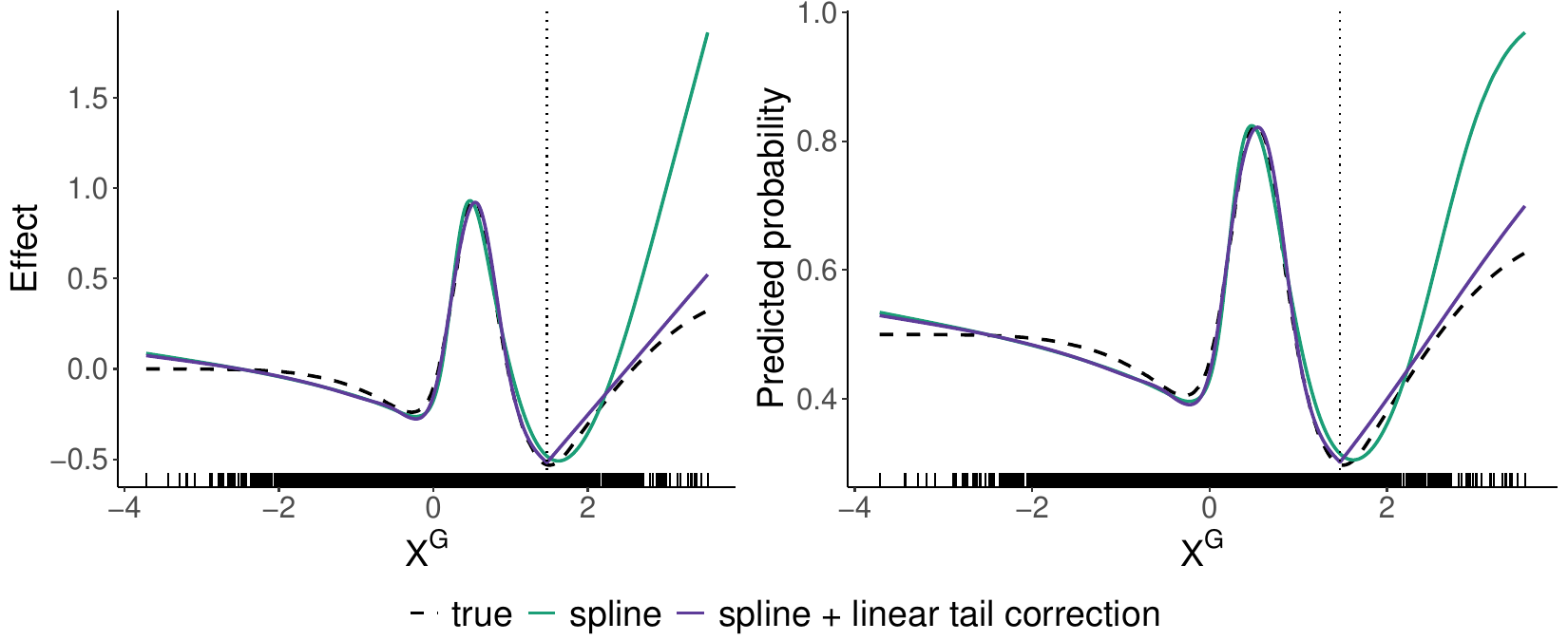}\vspace{5mm}
    \includegraphics[width=\linewidth]{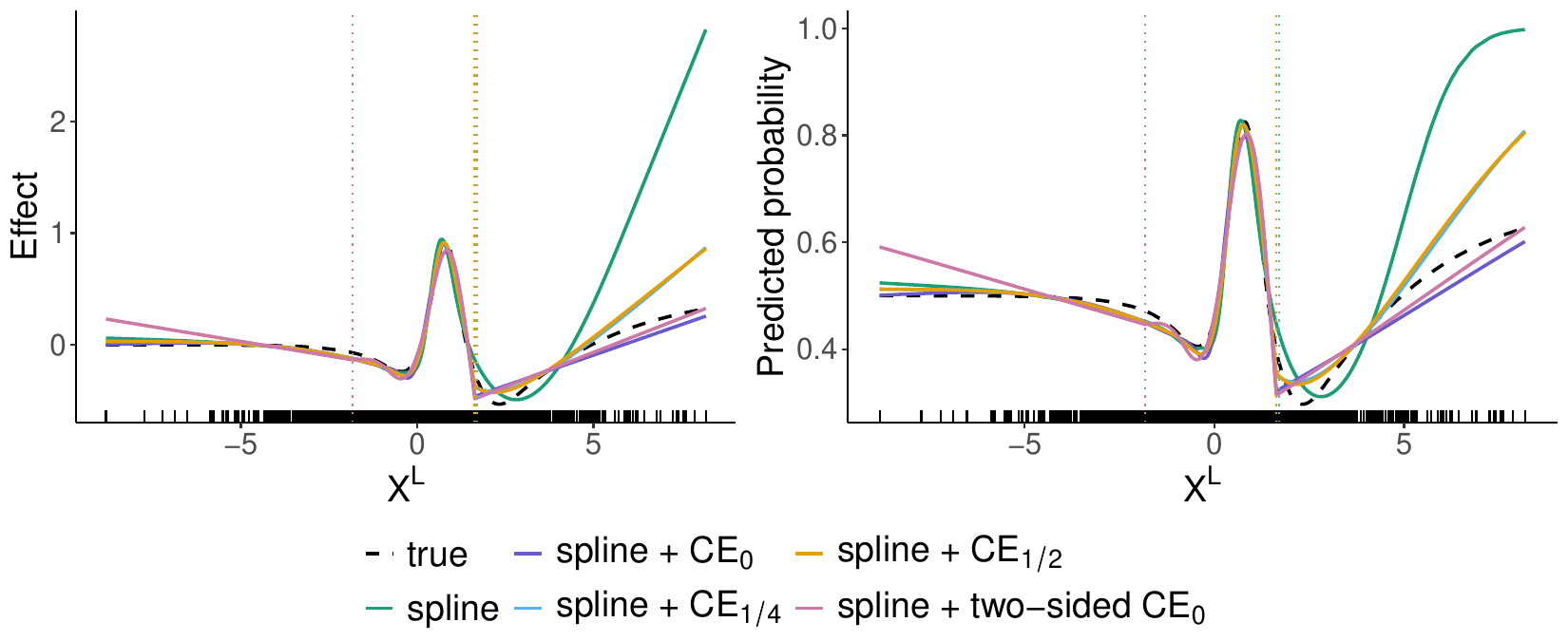}\vspace{5mm}
    \includegraphics[width=\linewidth]{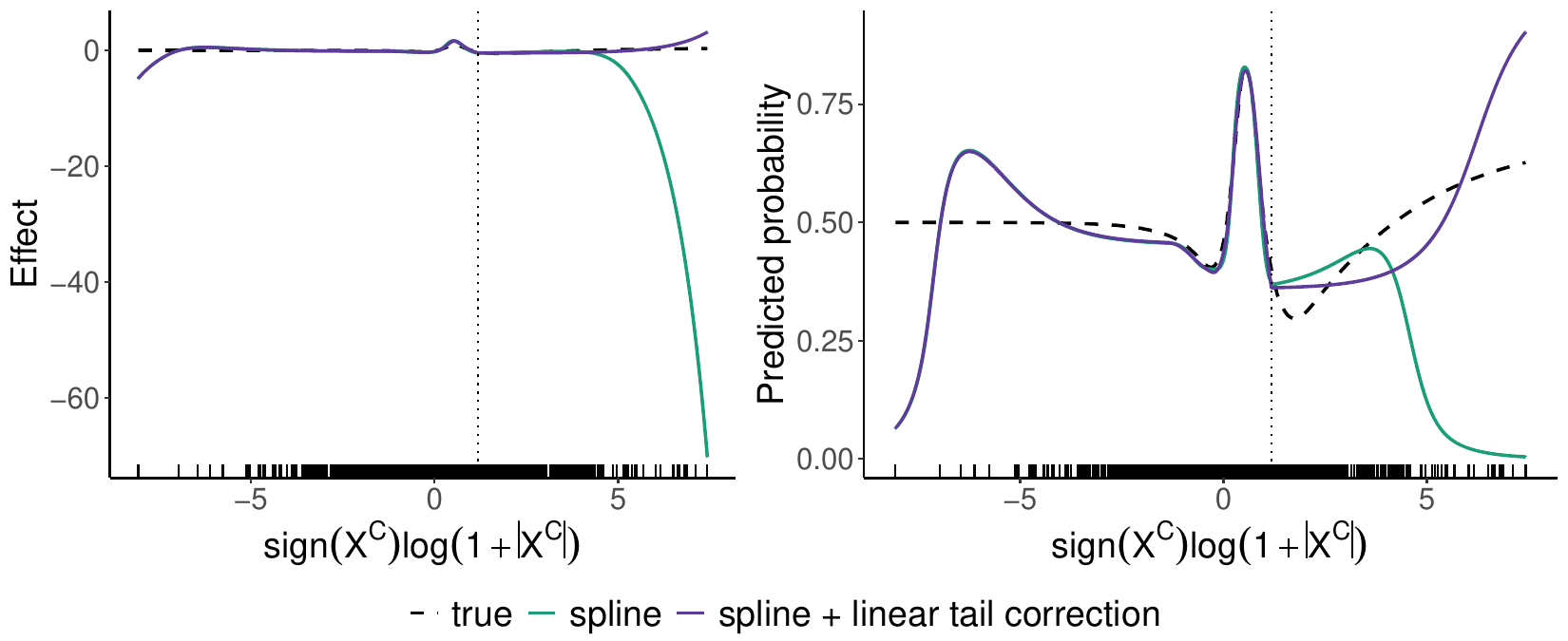}
    \caption{\footnotesize Example of fitted binary regression models under different frameworks. Left: true and estimated covariate effects evaluated on the test set. Right: predicted versus true conditional probabilities $\Pr(Y=1\mid X)$ for the test set. Dotted vertical lines mark the threshold where linearity is imposed for each model.}
    \label{fig:sim-binaryCE}
\end{figure}
Figure~\ref{fig:sim-binaryCE} compares, for the data shown in the first row of Figure~\ref{fig:sim-bin}, the estimated covariate effects and predicted probabilities under different frameworks, with and without tail linearity imposed in the upper tail of the covariate. 
Across all scenarios, the model based solely on a spline produces predicted probabilities 
very different from the truth, whereas a tail correction beyond a high threshold consistently yields a better fit.

While Figure~\ref{fig:sim-binaryCE} illustrates a single example, Table~\ref{tab:scores-binCE} summarizes out-of-sample performance across all 100 simulated datasets. We compare logarithmic score, AUC, AUPRC, their weighted counterparts for observations with extreme covariate values in the test set, and the Akaike information criterion (AIC). 
According to AIC, the simplest (baseline) probit model with a spline on the original covariate scale  provides on average the best fit. However, pre-transforming the covariate and combining spline and linear tail correction improves predictive performance across all other metrics, especially the weighted scores. Models on Laplace margins with tail correction motivated by the conditional extreme framework perform best for extreme covariate values, while remaining overall competitive with the corresponding model on Gaussian margins. 
Models on Cauchy margins are more sensitive to outliers, but the linear tail component still improves performance over the baseline model for extreme covariates. 

\begin{table}[h]
\centering
\resizebox{\textwidth}{!}{
\begin{tabular}{lccccccc}
\hline
\multicolumn{1}{c}{}       & \multicolumn{4}{c}{\textbf{On all the observations}}                & \multicolumn{3}{c}{\textbf{Only on the extremes}} \\ \cmidrule(lr){2-5} \cmidrule(l){6-8}
\textbf{Model}             & \textbf{\small LogS} & \textbf{\small 1-AUC} & \textbf{\small 1-AUPRC} & \textbf{\small AIC}    & \textbf{\small LogS$w$} & \textbf{\small 1-AUC$w$} & \textbf{\small 1-AUPRC$w$}\\ \hline
Probit GAM for $Y\mid X$, spline & 7.01 & 6.46 & 6.51 & \textbf{4.12} & 8.42 & 7.00 & 6.72\\  \hline
Probit GAM for $Y\mid X^{(G)}$, spline            &    6.59 & 6.20 & 6.37 & 4.80 & 7.72 & 6.20 & 6.24 \\
\quad + linear tail correction   &   4.08 & 5.86 & 5.96 & 5.54 & \textbf{3.22} & 5.50 & 5.79 \\ 
\quad + two-sided linear tail correction & \textbf{3.18} & 3.77 & 3.79 & 6.79 & 4.14 & 5.71 & 5.92 \\
\hline
Probit GAM for $Y\mid X^{(L)}$, spline                      & 7.65 & 6.62 & 6.85 & 5.30 & 7.97 & 6.67 & 6.64      \\
\quad + CE $\gamma = 0$   & 5.90 & 6.58 & 6.49 & 6.34 & 4.04 & 5.54 & 5.55    \\
\quad + CE $\gamma = 1/4$ & 5.60 & 5.66 & 5.13 & 5.70 & 5.10 & 4.39 & 4.34          \\
\quad + CE $\gamma = 1/2$ & 5.38 & 5.66 & 5.25 & 5.82 & 4.82 & \textbf{4.29} & \textbf{4.29} \\
\quad + two-sided CE $\gamma = 0$ &     3.24 & \textbf{3.12} & \textbf{3.53} & 5.92 & 4.51 & 5.77 & 5.84 \\ \hline
Cauchit GAM for $Y\mid X^{(C)}$, spline                &  10.0 & 9.23 & 9.29 & 8.38 & 9.83 & 8.88 & 8.58  \\
\quad + linear tail correction        &     7.35 & 6.85 & 6.83 & 7.28 & 6.23 & 6.05 & 6.10\\ \hline

\end{tabular}}
\caption{\footnotesize Average ranks of out-of-sample performance measures for the binary-response simulations: logarithmic score (LogS), complements of the area under the ROC curve (AUC) and precision--recall curve (AUPRC), computed over the full test set and over the 500 observations with the largest values of $X$. Akaike information criterion (AIC) is also reported. The best value for each measure is in bold.
}\label{tab:scores-binCE}
\end{table}

\subsection{Simulations for continuous response}
\begin{figure}[h]
    \centering
    \includegraphics[width=\linewidth]{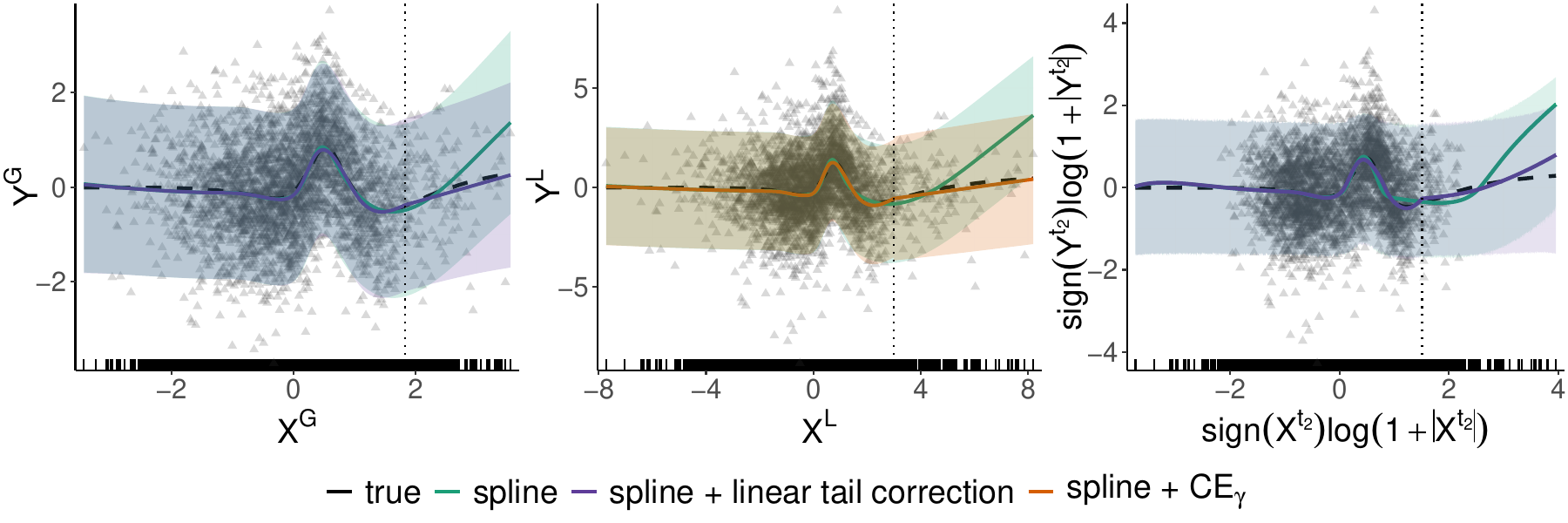}
    \caption{\footnotesize Different fitted models in the continuous-response simulation. Conditional mean and 95\% prediction interval on the test set are compared with the true data-generating function. The dotted vertical line marks the threshold where linearity is imposed. Left: standard Gaussian margins; center: standard Laplace margins; right: Student's $t_2$ margins transformed for improved readability.}
    \label{fig:sim-contCE}
\end{figure}

We use the same simulation scenarios as above, now taking $Y=\tilde{Y}$. 
The comparison between the models arising from different frameworks is shown in Figure~\ref{fig:sim-contCE} for a single dataset. 
The results mirror the ones obtained for the binary response, with a even better performance of the model that combines a spline for the bulk with a linear structure in the tail of the covariate.

Table~\ref{tab:scores-contCE} reports the average ranks across the different performance measures, including a standard Gaussian GAM fitted on the original scale. Since AIC values are not comparable across models with different response transformations, AIC rankings should only be interpreted within the same marginal scale of $Y$.
Overall, transforming the data to appropriate marginal scales and incorporating a linear tail correction improves predictive performance, also relative to the model fitted on the original scale. The model on Student-$t$ margins with linear tail correction performs particularly well, but the specification on Laplace margins provides the best performance for extreme covariate values.

\begin{table}[h]
\centering
\resizebox{\textwidth}{!}{
\begin{tabular}{lccccc}
\hline
\multicolumn{1}{c}{} & \multicolumn{3}{c}{\textbf{On all the observations}}         & \multicolumn{2}{c}{\textbf{Only on the extremes}} \\ \cmidrule(lr){2-4} \cmidrule(lr){5-6} 
\textbf{Model}       & \textbf{LogS} & \textbf{CRPS} & \textbf{
\textbf{AIC}
} &
\textbf{LogS$_w$}     & \textbf{CRPS$_w$}     \\ \hline
Normal GAM for $Y \mid X$, spline & 5.19 & 5.23 & - & 4.56 & 4.57  \\ \hline
Normal GAM for $Y^{(G)} \mid X^{(G)}$, spline             & 3.40 & 4.30 & \textbf{1.42} & 4.64 & 4.94   \\
\quad + linear tail correction        & 3.06 & 3.23 & 1.58 & 3.99 & 3.90       \\ \hline
Normal GAM for $Y^{(L)} \mid X^{(L)}$, spline            &  5.39 & 4.57 & \textbf{1.46} & 4.26 & 4.25             \\
\quad + linear tail correction        & 4.53 & 3.30 & 1.54 & \textbf{2.96} & \textbf{2.62}    \\ \hline
Scaled-$t$ GAM for $Y^{(t_2)} \mid X^{(t_2)}$, spline             &     3.56 & 4.29 & 1.58 & 4.18 & 4.48               \\
\quad + linear tail correction        & \textbf{2.87} & \textbf{3.08} & \textbf{1.42} & 3.41 & 3.24    \\ \hline
\end{tabular}}
\caption{\footnotesize Average ranks of out-of-sample performance measures for the continuous-response simulations: logarithmic score (LogS) and CRPS, computed over the full test set and over the 500 observations with the largest values of $X$. Akaike information criterion (AIC) is also reported, but should only be compared across models with the same marginal scale for $Y$.
The best values are in bold.
}\label{tab:scores-contCE}
\end{table}

\section{Application to wildfires in Europe}\label{sec:applications}

The prediction of extreme wildfires is a critical issue for European forest ecosystems with increasing occurrence numbers and sizes observed in many regions against the backdrop of climate change.  In future scenarios, many of the environmental and climatic drivers of wildfires further increase in magnitude, so that covariate extrapolation naturally applies to wildfire prediction \citep{pimont_lengthening_2022,ruffault:hal-05121416}. We analyze a dataset of wildfires in Europe compiled by \cite{ruffault:hal-05121416}. A fire occurs if its associated burnt area is greater than 100 ha (which are the fires for which reliable data is available), and, following \cite{ruffault:hal-05121416}, it is considered extreme if it exceeds 5000 ha. The dataset contains more than 150\,000 observations of wildfires and control points without wildfires from 2008 to 2023, for a total of 7\,276 fires (4.6\% of the total records) and 136 extreme fires (0.1\% of the total records and 1.9\% of the fires).

\subsection{Wildfires getting extreme}
We focus on modeling the probability of a wildfire escalating to an extreme wildfire. Let our response variable be $Y=1$ if a fire is extreme and $Y=0$ if a fire is not extreme. We use the same covariates selected for this problem by \cite{ruffault:hal-05121416}: daily mean vapor pressure deficit (VPD), daily mean wind speed (WS), daily mean continuous Haines index (CHI), broadleaf fraction (broadleaf) and percentage of wild land area (wildland). We fit a traditional probit model with splines for these five covariates, as well as a probit models with the covariates on Laplace margins, using splines and also imposing a linear or constant extrapolation in the tail. In particular, a linear effect in the tail was found to be significant only for the Haines index, thus for the other variables we can use simply a spline on their whole range or impose a constant effect in the tail. Since the sample size and the number of extreme fires are not very large, we implement a cross-validation procedure, leaving out one year at a time, and at the end obtaining predictions of the probability of escalating to an extreme for all the 7\,276 fires, that can be used to compute measures of model performance.

\begin{figure}[h]
    \centering
    \includegraphics[width=\linewidth]{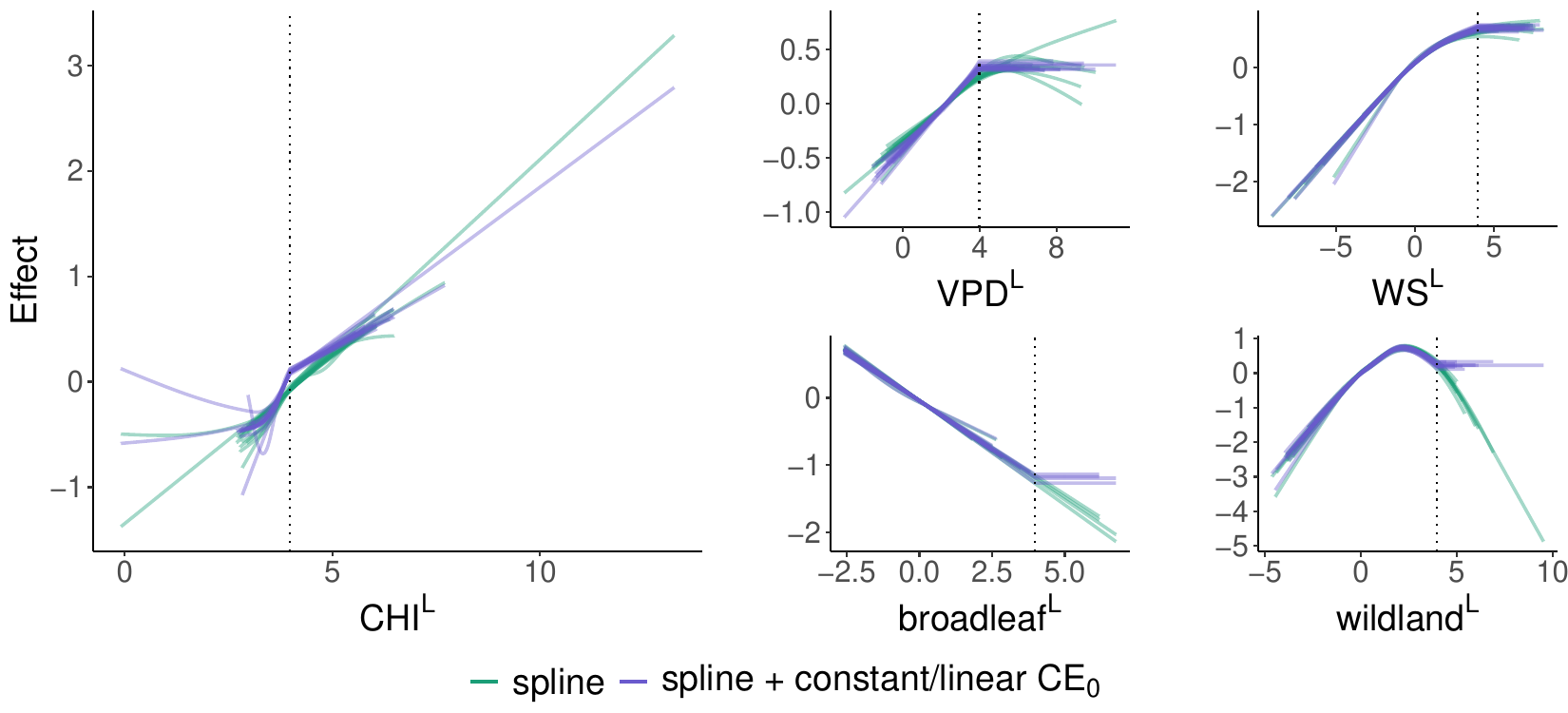}
    \caption{\footnotesize Estimated covariate effects (on standard Laplace margins) on the probability of escalation to a large wildfire across years. Models use either a spline or a spline below a high threshold and a conditional extreme model (linear or constant) above it. The dotted vertical line marks the threshold.}
    \label{fig:effects-EWE}
\end{figure}

Table~\ref{tab:scores-EWE} compares different models based on some measures of model performance for binary response. These scores highlight the improvement of the models with covariates on Laplace margins, and, focusing on the extremes, especially the one that combines splines in the bulk of the covariates with a linear model for the tail of the Haines index and a constant model for the tail of the other covariates. Figure~\ref{fig:effects-EWE} shows a comparison of the partial effects across years of the covariates on Laplace margins based on this model and the one with just splines. 

\begin{table}[h]
        \centering
        \small
        \renewcommand\arraystretch{1.2}
        \resizebox{\textwidth}{!}{
        \begin{tabular}{lccccccc}
        \hline
        & \multicolumn{4}{c}{\textbf{On all the observations}}  & \multicolumn{3}{c}{\textbf{Only on the extremes}} \\ \cmidrule(lr){2-5} \cmidrule(l){6-8}
       \textbf{Model} & \textbf{LogS} & \textbf{\footnotesize 1-AUC} & \textbf{\footnotesize 1-AUPRC}  & \textbf{\footnotesize AIC}  & \textbf{LogS$w$} & \textbf{\footnotesize 1-AUC$w$} & \textbf{\footnotesize 1-AUPRC$w$} \\ \hline
     Probit GAM for $Y \mid \bm X$, splines & 0.081 & 0.200 & 0.909 & 1093.4 & 0.196 & 0.339 & 0.895 \\ 
       Probit GAM for $Y\mid \bm X^{(L)}$, splines & 0.079            & 0.171 & 0.905 & 1070.4 & 0.190                      & 0.298       & 0.871     \\
     \quad + linear tail correction & \textbf{0.078}                             & 0.171 & \textbf{0.899}  & 1073.0        & 0.189                    & 0.288 & \textbf{0.869}\\
   \quad + linear/constant tail correction & \textbf{0.078}                & \textbf{0.169} & \textbf{0.899}  & \textbf{1069.4}       & \textbf{0.185}            & \textbf{0.284} & 0.870\\
        \hline
        \end{tabular}}
        \caption{\footnotesize Out-of-sample performance for predicting the occurrence of extreme wildfires: logarithmic score (LogS), complements of the area under the ROC curve (AUC) and precision--recall curve (AUPRC), computed over the full data set. Threshold-weighted LogS, AUC, and AUPRC are also reported, with weights emphasizing the 500 observations having the largest maximum covariate value on Laplace margins. Akaike information criterion (AIC) is also reported. The best values are in bold.}\label{tab:scores-EWE}
\end{table}

\subsection{Burnt area}
Another important goal in wildfire studies is to predict the size of wildfires, i.e., the burnt area. The response variable $Y$ is now continuous and taking values greater or equal to 100 ha. We consider only the burnt area of the 7\,276 occurred fires. We include as covariates all the variables selected by \cite{ruffault:hal-05121416} for the analysis of wildfire occurrences and escalation to extreme wildfires, meaning that to ones mentioned in the previous section we add daily drought code (DC), aridity index (AI), and topographic roughness index (TRI). Duff moisture code (DMC) was at first included but proved to have no effect on the burnt area. 

Burnt area data are notoriously heavy-tailed, due to the occurrence of many small fires and few extremely large fires. Some usual modeling options for this kind of data are log-normal GLM or GAM, and Gamma GLM or GAM with log link function. An alternative model motivated by extreme value theory is the GPD for threshold exceedances with scale (and sometimes shape) parameter informed by the data. A flexible extension of this model is the EGPD \citep{naveau2016modeling}, that allows to model both the bulk and the tail 
while retaining the same asymptotic tail behavior of the GPD. We compare these more traditional approaches to the proposed model based on the conditional extreme framework. This means that response and covariates are transformed to Laplace margins, and a normal location–scale GAM is then fitted, using a linear structure in the tail of the covariates. 

The fitted models are compared in Table~\ref{tab:scores-burntarea} using scoring rules, their weighted counterparts, and AIC (we recall that AIC is only comparable across models with the same response variable). We also extract from each model the predicted probability of a fire becoming extreme, i.e., exceeding 5\,000 ha, and evaluate model performance in terms of classification of extreme events based on the AUPRC. 
The Gamma model (whose estimated shape is 0.24) is the best one at this task, even when the covariates are extreme. The GPD, EGPD and normal models on Laplace margins have the best predictive performance overall, with the normal ones performing best for extreme covariates.

\begin{table}[h]
        \centering
        \small
        \renewcommand\arraystretch{1.2}
        \setlength{\tabcolsep}{3.5pt}
        \resizebox{1.01\textwidth}{!}{
        \begin{tabular}{lccccccc}
        \hline
        & \multicolumn{4}{c}{\textbf{On all the observations}}  & \multicolumn{3}{c}{\textbf{Only on the extremes}} \\ \cmidrule(lr){2-5} \cmidrule(l){6-8}
       \textbf{Model} & \textbf{\footnotesize LogS} & \textbf{\footnotesize CRPS}  & \textbf{\footnotesize 1-AUPRC} & \textbf{\footnotesize AIC}  & \textbf{\footnotesize LogS$w$} & \textbf{\footnotesize CRPS$w$} & \textbf{\footnotesize 1-AUPRC$w$} \\ \hline
      Gamma GAM with log link for $(Y-100)/100$ &  7.68 & 612.6 & \textbf{0.866} & 27147.2 &  7.95 &  955.9 & \textbf{0.866}\\ 
       Log-normal GAM for $Y-100$ &    7.39         & 693.2 & 0.900 & 27582.9 & 7.74 &  1066.3 & 0.905                        \\
       GPD GAM for $Y-100$ & \textbf{7.12} & 577.2 & 0.881 & 77157.1 &  7.50 & 920.7 & 0.889\\
       EGPD GAM for $Y-100$ & 7.17 & 577.3 & 0.879 & 76705.4 &  7.54 & 921.7 & 0.882\\
    Normal GAM for $Y^{(L)}$ &     7.15  & \textbf{574.1} & 0.879 & 20831.3  &     7.49   &         \textbf{914.1}    & 0.894    \\
    Normal location-scale GAM for $Y^{(L)}$ + linearity &            7.14     & 575.6 & 0.889 &  20823.1 &  \textbf{7.48}     & 915.3    & 0.899       \\
        \hline
        \end{tabular}}
        \caption{\footnotesize Out-of-sample performance for predicting wildfire sizes: logarithmic score (LogS), CRPS, complement of the precision--recall curve (AUPRC), computed over the full data set. Threshold-weighted LogS, CRPS, and AUPRC are also reported, with weights emphasizing the 500 observations having the largest maximum covariate value on Laplace margins. Akaike information criterion (AIC) is also reported. The best values are in bold.}\label{tab:scores-burntarea}
\end{table}

\begin{figure}[h]
    \centering
    \includegraphics[width=\linewidth]{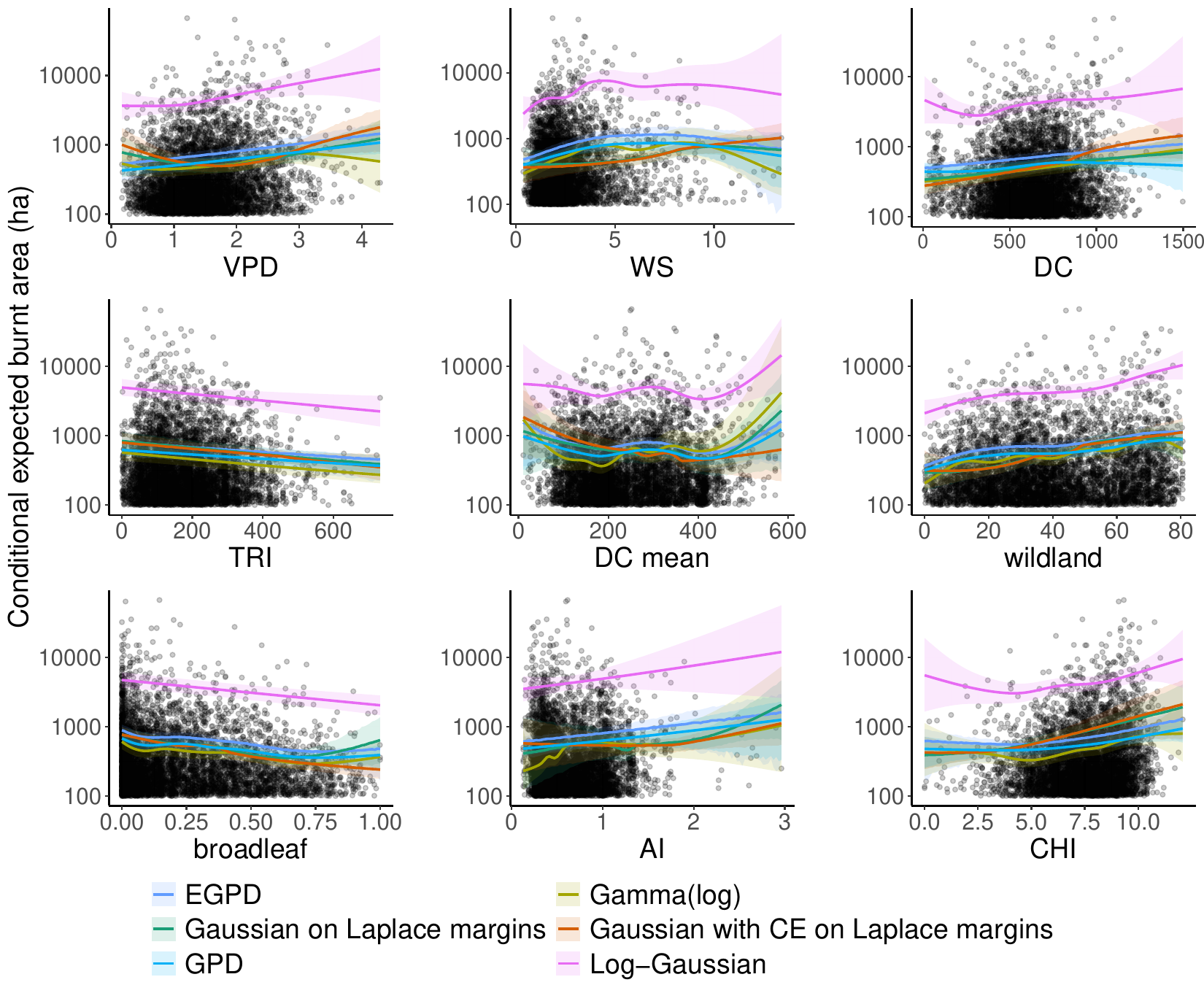}
    \caption{\footnotesize Effect of each covariate under different models on the expected fire size (on the $\log_{10}$ scale), conditional on all other covariates being fixed at their median. Shaded areas represent 95\% confidence intervals, and the point cloud shows the relationship between each covariate and the $\log_{10}$ burnt area in the test set.}
    \label{fig:fires-cont}
\end{figure}

The predicted burnt area from each model is displayed in Figure~\ref{fig:fires-cont} as a function of each covariate when keeping the other ones constant. The plots highlight that the log-normal model is not working well. 
The other models behave more similarly, with differences mainly in the right tail of covariates. It is possible to see the imposed linear or constant behavior for the tail-corrected model.

\section{Discussion}\label{sec:discussion}
Future covariates often include more extreme conditions than those observed before, with weather variables in climate-change applications being a prominent example.  This mismatch between training and prediction data can severely degrade performance of predictive models such as GAMs, particularly in the tails of the covariate distributions. By leveraging multivariate tail representations, we recommend combinations of marginal transformations, response distributions---both binary and continuous---and link functions that are theoretically supported by multivariate extreme value theory.

GAM implementations such as those in the widely used \texttt{mgcv} package of the \emph{R} statistical software already put linearity constraints at the extreme values of the covariate training sample as a default choice. In addition, we propose using linear extrapolation after first pretransforming the covariates and response to specific marginal scales. Moreover, we allow for transitions to linearity already within the range of the covariate training sample by considering the transition threshold as  a hyperparameter for which several values can be checked to select the best one. 
Our results indicate that model performance can improve when covariates and response (if continuous) are pre-transformed to marginal scales motivated by theoretical results. Additional gains can be achieved by imposing a linear structure in the tails of the covariates prior to the linearity constraints at the boundary implemented in \texttt{mgcv}. These improvements are not universal across all settings and datasets; however, the proposed approach proves to be a useful tool in many cases. We therefore recommend considering its use as part of the modeling strategy, particularly when tail behavior is of interest.

Constant extrapolation of covariate effects, as imposed by many standard machine learning models, is inappropriate in many applications. Nevertheless, it can be the appropriate choice in cases of saturation effects where increasing a covariate beyond a large threshold (or decreasing it below a small threshold) does not further modify the behavior of the response. Therefore, including a model with constant extrapolation in the list of models to select from is useful when the modeling framework does not naturally impose or tend towards constant extrapolation. 

We targeted moderate extrapolation of covariates when moving slightly beyond the observed range of values. Statistical extrapolation of covariate effects to very extreme quantiles will always be associated with very high uncertainties, except for cases where the true relationship between response and covariate actually is linear. More reliable extrapolation would require incorporating  expert knowledge about known physical mechanisms, 
which is beyond the scope of this work.  

Alternative approaches enabling linear extrapolation beyond the observed range of the covariates have been proposed. One example is local linear forests \citep{friedberg2020local}, which combine random forests and local linear regression. However, their extension to binary response settings is not straightforward, as they do not accommodate the specification of a link function. A detailed comparison with this method is not in the scope of this work, as our main focus is on generalized additive models. 

Regarding future work, geometric extremes \citep{nolde2022linking} could be interesting for covariate extrapolation under exponential tail assumptions beyond more classical multivariate extreme value models, specifically in the GAM setting \citep{murphy2025modelling}. However,  understanding the implications of this framework for predictive distributions would first require further theoretical developments.

Beyond the GAM framework, our approach could be naturally extended to neural networks, particularly those using the standard Rectified Linear Unit (ReLU) activation function \citep[e.g.,][]{nair2010rectified}, or other activation functions having a similar, linear behavior for high input values. ReLU-based architectures induce linearity outside the range of the training data, thus implicitly allowing linear extrapolation. Therefore, the theoretical motivation and the marginal pre-transformations developed here would be applicable, but with the added flexibility of neural networks.

\newpage
\appendix

\section{Marginal transformations to standardized scales}\label{app:marginal}

\vc{Given a random vector $\bm W = (W_1,\dots,W_m)\sim F_{\bm W}$ with continuous marginal distributions $F_{W_j}$, 
a common choice is 
to transform each $W_j$ to a common marginal scale. We discuss here some marginal distributions of the transformed $W_j^\star$.
}

\subsection*{Exponential-tailed distributions}
\vc{Many marginal distributions that are common in multivariate extreme value theory are part of the class of exponential-tailed distributions, meaning that
$\text{Pr}(W_j^\star > t+x)/\text{Pr}(W_j^\star > t) \rightarrow \exp(-\alpha x)$ for $x>0$ as $t\rightarrow\infty$, with rate parameter $\alpha \geq 0$.
More specifically, for large $x$, one usually considers $W_j^\star$ with tail probabilities 
\begin{equation*}
    \text{Pr}(W_j^{(ET)}>x)\sim C e^{-cx},
\end{equation*}
for some constants $C,c>0$, and we use the $^{(ET)}$-superscript to refer to such margins. Some examples with $c=1$ are:}
\begin{itemize}
    \item \vc{Standard exponential distribution: $\text{Pr}(W_j^{(E)}>x)=e^{- x}$, $x>0$. 
    \item Standard Laplace distribution: $\text{Pr}(W_j^{(L)}>x)=e^{- x}/2$, $x \in \mathbb{R}$.
    \item Modified standard Laplace distribution, that is introduced below: $\text{Pr}(W_j^{(\tilde{L})}>x)=e^{- x}e/4$, $x \in \mathbb{R}$.}
\end{itemize}

\subsubsection*{Modified Laplace distribution}\label{sec:laplace-alt}
The Laplace distribution puts a lot of mass around 0, and when fitting a spline on this marginal scale the resulting smooth partial effect could be very wiggly and not flexible enough for values close to 0.
It is possible that this behavior for covariate values around 0 leads to worse prediction performance for the non-extreme covariates when using the Laplace distribution.
Thus we want to use a marginal scale that
still has exponential tails on both sides, but that has a density that is less spiky around 0, closer to the flat density of the normal distribution close to its mean. This alternative distribution has probability density function defined as
\begin{equation}\label{eq:uniflaplace}
    f(x) = \begin{cases}
        1/4 & x \in [-1,1],\\
        \exp\{-(\vert x \vert -1)\}/4 & \vert x\vert >1.
    \end{cases}
\end{equation}
The cumulative distribution function is given by
\begin{equation*}
    F(x) = \begin{cases}
        \exp(x+1)/4 & x<-1,\\
        1/2 + x/4 & x \in [-1,1],\\
        1 - \exp\{-(x-1)\}/4 &  x >1,
    \end{cases}
\end{equation*}
and the quantile function is
\begin{equation*}
     Q(p) = \begin{cases}
        \log(4p)-1 & p<0.25,\\
        4p-2 & p \in [0.25,0.75],\\
        1-\log(4(1-p)) &  p >0.75.
    \end{cases}
\end{equation*}
\begin{figure}[t]
    \centering
    \includegraphics[width=0.87\linewidth]{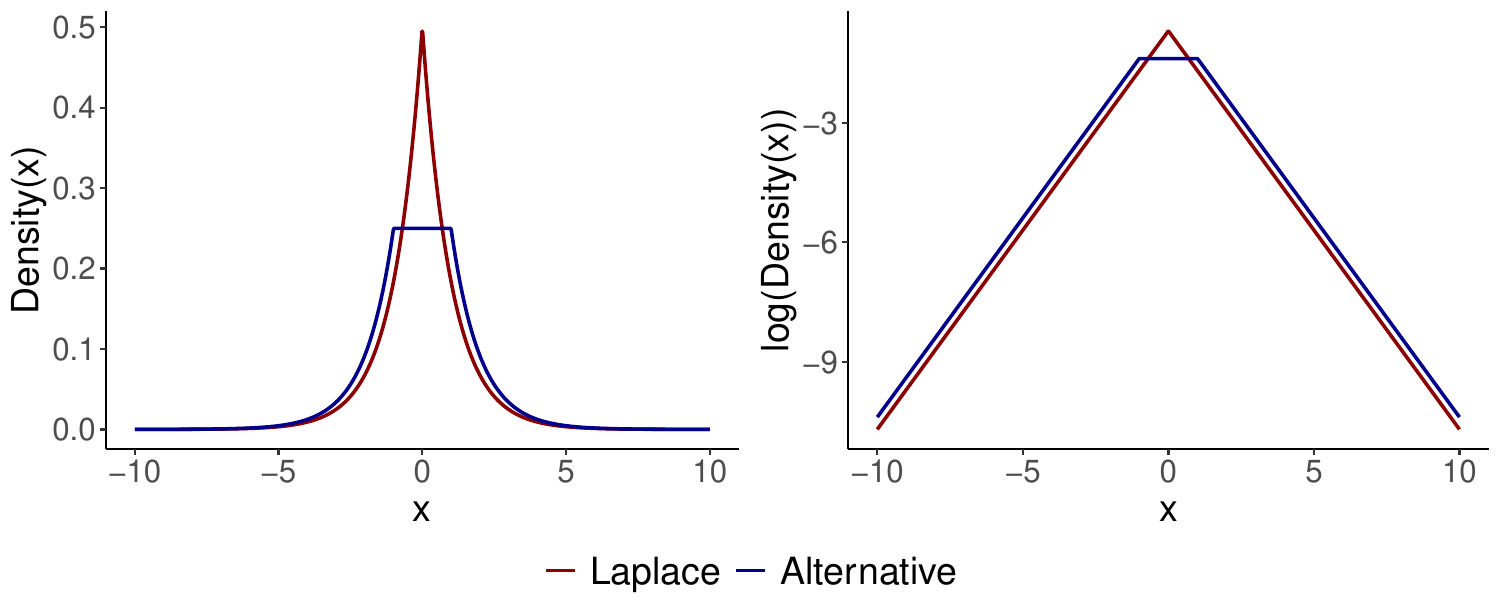}
    \caption{\footnotesize Comparison between standard Laplace and modified Laplace  distribution. Left: probability density functions. Right: logarithm of the probability density functions.}
    \label{fig:laplace-alt}
\end{figure}
Figure~\ref{fig:laplace-alt} shows the difference between this distribution and a standard Laplace.

\subsection*{Regularly varying and Pareto-tailed distributions}

\vc{Univariate regular variation (at infinity) of a standardized random variable  is defined through regularly varying survival functions:
\begin{equation*}
    \text{Pr}(W^{(RV)}_j > x) \sim  \ell(x)x^{-\alpha},
\end{equation*}
where $\alpha>0$ is the regular variation index and $\ell$ is a slowly varying function, for each marginal variable $W^{(RV)}_j$. Distributions with this tail behavior are sometimes also called power-law distributions. This class characterizes the marginal distributions relevant for multivariate regular variation. In practice, we use distributions where $\ell(x)$ behaves like a constant $c>0$, and we call them Pareto-tailed distributions, indicated by superscript $^{(PT)}$:
\begin{equation*}
    \text{Pr}(W^{(PT)}_j > x) \sim  cx^{-\alpha}.
\end{equation*}
}
\vc{
Several common marginal transformations belong to the Pareto-tailed subclass with $\alpha=1$ and $c=1$, including }
\begin{itemize}
    \item \vc{The standard Pareto distribution: $\text{Pr}(W^{(P)}_j > x)=x^{-\alpha}$, $x \geq 1$. 
    \item The standard Fréchet distribution: $\text{Pr}(W_j^{(F)} > x) = 1-\exp(-x^{-1}) \sim x^{-1}$, $x>0$.
    \item The standard log-Laplace distribution: if $W_j^{(L)}$ is standard Laplace, then $W_j^{(logL)} = 2e^{W_j^{(L)}}$
    has survival function $\text{Pr}(W_j^{(logL)} > x) = x^{-1}$, $x\geq 1$.}
\end{itemize}
\vc{
Some other useful Pareto-tailed distributions with different values for $c$ and/or $\alpha$  are as follows:
\begin{itemize}
    \item The Cauchy distribution, with $\alpha=1$ and $c=1/\pi$.
    \item The Student's $t$ distribution with $\nu$ degrees of freedom, with $\alpha=\nu$ and $c=\frac{\Gamma\left((\nu+1)/2\right)}
    {\sqrt{\nu \pi}\Gamma(\nu/2)}
    \nu^{(\nu-1)/2}$.
    \item The modified log-Laplace distribution: if $W_j^{(\tilde{L})}$ follows the modified-Laplace distribution of \eqref{eq:uniflaplace}, then $W_j^{(log\tilde{L})} = e^{W_j^{(\tilde{L})}}$ has regularly varying survival function with $\alpha=1, c=e/4$.  
\end{itemize}
If $W_j^{(PT)}>0$ is Pareto-tailed, then, equivalently, $\log W_j^{(PT)}$ is exponential-tailed.}

\section{Details of simulation scenarios}\label{app:scenarios}
The plots shown in Figure~\ref{fig:scenarios} are based on data simulated under four different scenarios, each one corresponding to one row of the figure. 
\begin{enumerate}
    \item $X \sim \text{Exp}(1)$, $\varepsilon \sim \text{N}(0,1)$, $Y=X/2+\varepsilon$.
    \item $X \sim \text{Pareto}(2)$, $\varepsilon \sim \text{N}(0,1)$, $Y=X/2+\varepsilon$.
    \item $X \sim \text{Pareto}(2)$, $\varepsilon \sim \text{t}_2$, $Y=2X+3\varepsilon$.
    \item $X \sim \text{Exp}(1)$, $\varepsilon \sim \text{Laplace}(0,2)$, $Y=X/2+\varepsilon$.
\end{enumerate}
Then $X$ is transformed to different marginal scales:
\begin{itemize}
    \item Uniform margins: $X^{(U)} = F_X(X)$,
    \item Pareto margins with shape 2: $X^{(P)} = [1/\{1-F_X(X)\}]^{1/2}$,
    \item Standard exponential margins (if not already): $X^{(E)} = F^{-1}_{\text{Exp}}\{\hat{F}_X(X)\}$, where $F^{-1}_{\text{Exp}}$ is the quantile function of the exponential distribution with rate 1, and $\hat{F}$ denotes the empirical distribution function.
\end{itemize}
$Y$ is also transformed to standard exponential margins by computing $Y^{(E)} = F^{-1}_{\text{Exp}}\{\hat{F}_Y(Y)\}$.

Results based on these simulations show that a Gaussian GAM generally well captures the mean relationship between covariate and response. However, a Student-$t$ GAM can provide an equally good or better fit, especially when the covariate is heavy tailed, while also yielding lower uncertainty around the estimate.

\section{Details about the multivariate Gaussian model}\label{app:mv-gaussian}
Assume that $(Y^{(G)},\boldsymbol X^{(G)})$ has a joint multivariate normal distribution with standard normal margins. Then, the classical Gaussian linear regression assumes that the conditional distribution of $Y^{(G)}$ given $\boldsymbol X^{(G)}$ is normal with linear mean function $\sum_{j=1}^{d} \beta_j X_j^{(G)}$.
By a standard property of the multivariate normal distribution \citep[see, e.g.,][Sec.~2.5]{Anderson2003}, if we knew the covariance matrix of $(Y^{(G)},\boldsymbol X^{(G)})$, we could exactly calculate the regression coefficients $\beta_j$, and define the point predictor as $\hat{Y}(\bm X^{(G)}) = \sum_{j=1}^{d} \beta_j X_j^{(G)}$. The pair $(Y^{(G)},\hat Y(\boldsymbol X^{(G)}))$ is therefore bivariate Gaussian. Excluding the case of perfect prediction (i.e., $\sigma_\varepsilon=0$), the coefficient of extreme point prediction satisfies $\chi(Y^{(G)},\hat{Y}(\bm X^{(G)}))=0$, because any bivariate Gaussian distribution with correlation strictly less than one is asymptotically independent \citep{sibuya1960bivariate,ledford1996statistics}. Only in the degenerate case with perfect prediction, $\chi(Y^{(G)},\hat{Y}(\bm X^{(G)}))=1$.

\section{Multivariate regular variation and risk functions}\label{app:mrv}

Assume that the random vector $\bm W\sim F_{\bm W}$ has standard Pareto margins. Then, $F_{\bm W}$ is multivariate regularly varying if
\begin{equation*}\label{eq:mrv}
    t (1-F_{\bm W}(t\bm w)) \rightarrow V(\bm w), \quad t \rightarrow \infty, \quad \bm w\ge0, \, \bm w \neq \bm 0,
\end{equation*}
where $V$ is a homogeneous dependence function satisfying $tV(t\boldsymbol{ x})=V(\boldsymbol{x})$, for all $t>0$ and $\boldsymbol x\geq \boldsymbol 0$.


Multivariate regular variation is the basis for defining $r$-Pareto processes as limits of processes when conditioning on  exceedance  of risk functionals; see the book chapter \citet{dombry2026pareto} for details. 
We here adapt the presentation of these limits to be useful for our regression setting.  
Consider a homogeneous risk function $r:[0,\infty)^d\rightarrow [0,\infty)$, continuous at the origin, where $r(t\boldsymbol x) = tr(\boldsymbol{x})$ for all $t>0$ and $\boldsymbol x\geq \boldsymbol 0$ and $r(\bm x_n)\rightarrow 0$ if $0\leq \min_{j=1}^d x_j\rightarrow 0$; for example, $r$ could be a norm. 
Then, under suitable regularity conditions, multivariate regular variation of $(\boldsymbol X,Y)$ ensures that exceedances above large values of $r(\boldsymbol X)$ admit the following limit:
\begin{equation*}
   \Pr\left( \frac{(\boldsymbol X, Y)}{r(\boldsymbol{X})}\in A \mid r(\boldsymbol X) \geq t\right) \rightarrow \mu_{r;V}(A), \quad t \rightarrow \infty, 
\end{equation*}
with a limit measure $\mu_{r;V}$ depending on $r$ and the dependence function $V$ of  $(\boldsymbol X, Y)$. 
We can choose $A=\{(u,v):v\le y\}$, and define a nonnegative residual vector $\varepsilon_r$, independent of $\bm X$, whose distribution $F_{\varepsilon_r}$ depends on $r$. In particular, $F_{\varepsilon_r}(y) = \mu_{r;V}(\{(u,v):v\le y\})$.
It follows that
\begin{equation*}
    \Pr\left( \frac{Y}{r(\boldsymbol{X})} \le y \mid r(\boldsymbol X) \geq t\right) = \Pr\left(Y \le r(\boldsymbol{X}) y \mid r(\boldsymbol X) \geq t\right) \rightarrow F_{\varepsilon_r}(y), \quad t \rightarrow \infty.
\end{equation*}

Then, multivariate regular variation ensures the following convergence, already stated in Equation~\ref{eq:Y-given-risk-exceedance}:
\begin{equation}
\left(\frac{Y}{r(\boldsymbol X)}\mid  r(\boldsymbol X) > t\right) \rightarrow r(\boldsymbol{X}) \varepsilon_r, \quad t\rightarrow \infty. 
\end{equation}

\section{Constraint on the $\gamma$-parameter in the conditional extremes model }\label{app:gamma-constraint}

Necessary constraints for the parameter $\gamma$ are usually given as $\gamma < 1$ in the formula \eqref{eq:bv-cond-extr} as explained by \citep{heffernan2004conditional}, but we here show that even more restrictive necessary constraints arise depending on the distribution of the error term $\varepsilon$ to avoid inconsistencies with the exponential marginal tails. These constraints arise from the requirement that the tail of the product term $(X_{j_0}^{(L)})^{\gamma} \varepsilon$ cannot be heavier than that of the standard Laplace distribution of $Y$. 
Indeed, \cite{arendarczyk2011asymptotics} show that the product of two independent random variables with Weibull-type tail decay also exhibits Weibull-type tail behavior. We recall that a distribution is of Weibull-type if its distribution function $F$ satisfies $F(v) = 1-\exp\{-v^p\ell(v)\}$, as $v \rightarrow \infty$, for $p$ a positive constant and $\ell(\cdot)$ slowly varying function at infinity \citep{hashorva2014tail}. For the standard Laplace distribution $p$ is equal to 1, and the following inequality must hold to ensure that the resulting tail of ${Y}^{(L)}\mid X_{j_0}^{(L)}$ is not heavier than Laplace: 
\begin{equation*}
    \frac{p_{\varepsilon}}{\gamma(p_{\varepsilon} + 1/\gamma)} \le 1,
\end{equation*}
where $p_{\varepsilon}$ denotes $p$ for the distribution of $\varepsilon$.
If $\varepsilon$ is assumed to follow a normal distribution, then necessarily $\gamma \in [0, 1/2]$. If $\varepsilon$ follows an exponential or Laplace distribution, then the constraint becomes $\gamma = 0$.
 To our knowledge, this necessary condition on $\gamma$ has not yet been stated in the literature on conditional extremes.

\section{Link functions for binary regression}\label{app:link}

In a GLM with binary response, the conditional probability of success depends on the covariates $\bm X$ through
$\text{Pr}(Y=1\mid \bm X)=g^{-1}\{\eta(\bm X)\}$, where $\eta(\bm X)$ is a linear combination of the covariates and $g$ is the link function.

\subsection*{Multivariate Gaussian model}
Under this model, the link function for a binary regression is the probit link $\Phi^{-1}$, since
\begin{align*}
    &\text{Pr}(\Tilde{Y}^{(G)} >u_Y\mid \boldsymbol{X}^{(G)}) =\text{Pr}\left(\varepsilon > u_Y - \sum_{j=1}^{d} \beta_jX_j^{(G)}\right)= \Phi \left( \frac{\sum_{j=1}^{d} \beta_jX_j^{(G)} - u_Y}{\sigma_{\varepsilon}}\right),
    \end{align*}
and thus $g^{-1}=\Phi.$

\subsection*{Conditional extreme model }
Under this framework,
\begin{align*}
    \text{Pr}\left(\Tilde{Y}^{(L)} > u_Y \mid X_{j_0}^{(L)} > u^{(L)}\right) 
    &= \text{Pr}\left(\varepsilon > u_Y \left(X_{j_0}^{(L)}\right)^{-\gamma}-\beta_{j_0}\left(X_{j_0}^{(L)}\right)^{1-\gamma}\right).
\end{align*}
The specific link function depends on the distribution of $\varepsilon$, as detailed below.

\subsubsection*{Normal error term}
If we assume $\varepsilon \sim \text{N}(0, \sigma^2_{\varepsilon})$, it follows that
\begin{align*}
    \text{Pr}\left(\varepsilon >u_Y^{(L)} \left(X_{j_0}^{(L)}\right)^{-\gamma}-\beta\left(X_{j_0}^{(L)}\right)^{1-\gamma}\right) &= \text{Pr}\left(\varepsilon < -u_Y^{(L)} \left(X_{j_0}^{(L)}\right)^{-\gamma}+\beta_{j_0}\left(X_{j_0}^{(L)}\right)^{1-\gamma}\right)\\
    &=\Phi\left\{\frac{-u_Y^{(L)} \left(X_{j_0}^{(L)}\right)^{-\gamma}+\beta_{j_0}\left(X_{j_0}^{(L)}\right)^{1-\gamma}}{\sigma_\varepsilon}\right\},
\end{align*}
and thus
\begin{equation*}
\Phi^{-1}\{\text{Pr}(Y = 1 \mid X_{j_0}^{(L)} > u^{(L)})\} = -\frac{u_Y^{(L)}}{\sigma_{\varepsilon}} \left(X_{j_0}^{(L)}\right)^{-\gamma}+\frac{\beta_{j_0}}{\sigma_{\varepsilon}}\left(X_{j_0}^{(L)}\right)^{1-\gamma} = \tilde{\beta}_1\left(X^{(L)}\right)^{-\gamma}+\tilde{\beta}_2\left(X^{(L)}\right)^{1-\gamma},
\end{equation*}
which is a probit model.


\subsubsection*{Laplace error term}
As mentioned in Section~\ref{sec:mv-extremes}, if we assume $\varepsilon \sim \text{Laplace}(0, \sigma_{\varepsilon})$, then $\gamma=0$ and the conditional extreme model simplifies to $\left({Y}^{(L)}\mid X_{j_0}^{(L)} > u^{(L)}\right) 
    = \beta_{j_0} X_{j_0}^{(L)} +  \varepsilon$. Thus, we obtain
\begin{align*}
    \text{Pr}(\varepsilon > u_Y -  \beta_{j_0} X_{j_0}^{(L)}) &= \begin{cases}
        1-\frac{1}{2} \exp \left(\frac{u_Y   -  \beta_{j_0} X_{j_0}^{(L)}}{\sigma_{\varepsilon}} \right), & u_Y   -  \beta_{j_0} X_{j_0}^{(L)} <0\\
        \frac{1}{2} \exp \left(\frac{-u_Y   +  \beta_{j_0} X_{j_0}^{(L)}}{\sigma_{\varepsilon}} \right), & u_Y   -  \beta_{j_0} X_{j_0}^{(L)} >0
    \end{cases}\\
    &\approx \begin{cases}
        1-\frac{1}{2} \exp \left(\frac{u_Y   -  \beta_{j_0} X_{j_0}^{(L)}}{\sigma_{\varepsilon}} \right), &  \beta_{j_0}>0\\
        \frac{1}{2} \exp \left(\frac{-u_Y   +  \beta_{j_0} X_{j_0}^{(L)}}{\sigma_{\varepsilon}} \right), &  \beta_{j_0} <0,
    \end{cases}
\end{align*}
where the approximation is as $X_{j_0}^{(L)} \rightarrow \infty$, which is our interest. The case $ \beta_{j_0}=0$ is not interesting, since it corresponds to a null effect of the covariate.
\begin{itemize}
    \item $ \beta_{j_0}>0$\\
    \begin{align*}
        \text{Pr}(Y = 0 \mid X_{j_0}^{(L)} > u^{(L)}) &= \frac{1}{2}\exp\left(- \frac{-u_Y   +  \beta_{j_0} X_{j_0}^{(L)}}{\sigma_{\varepsilon}} \right) \\ &=\exp\left(- \frac{-u_Y   +  \beta_{j_0} X_{j_0}^{(L)}+(\log 2)/\sigma_{\varepsilon}}{\sigma_{\varepsilon}} \right) \\ &= \exp(-w).
    \end{align*}
    Since $\exp(-w) \approx \exp(-w)/\{1+\exp(-w)\}$ for large $X_{j_0}^{(L)}$, we can approximate
    \begin{equation*} 
        \text{Pr}(Y = 1 \mid X_{j_0}^{(L)} > u^{(L)}) \approx \frac{\exp(w)}{1+\exp(w)} = \frac{\exp\left( \frac{-u_y   +  \beta_{j_0} X_{j_0}^{(L)}}{\sigma_{\varepsilon}} + \log 2 \right)}{1+ \exp\left( \frac{-u_y   +  \beta_{j_0} X_{j_0}^{(L)}}{\sigma_{\varepsilon}} + \log 2 \right)}.
    \end{equation*}
    Therefore,
    \begin{equation*}
    \text{logit}\{\text{Pr}(Y = 1 \mid X_{j_0}^{(L)} > u^{(L)})\} \approx \log2 -\frac{u_Y}{\sigma_\varepsilon} + \frac{ \beta_{j_0}}{\sigma_\varepsilon}X_{j_0}^{(L)} = \tilde{\beta}_1+\tilde{\beta}_2X_{j_0}^{(L)},
    \end{equation*}
    which is approximately a logistic regression model.
    \item $ \beta_{j_0}<0$\\
    \begin{align*}
        \text{Pr}(\varepsilon > u_Y   -  \beta_{j_0} X_{j_0}^{(L)}) &= \frac{1}{2} \exp \left(\frac{-u_Y   +  \beta_{j_0} X_{j_0}^{(L)}}{\sigma_{\varepsilon}} \right)\\
        &= \exp\left(- \frac{u_Y   -  \beta_{j_0} X_{j_0}^{(L)}+(\log 2)/\sigma_{\varepsilon}}{\sigma_{\varepsilon}} \right) \\ &= \exp(-z).
    \end{align*}
    For $X_{j_0}^{(L)} \rightarrow \infty$,
    \begin{equation*}
        \text{Pr}(\varepsilon > u_Y   -  \beta_{j_0} X_{j_0}^{(L)}) \approx \exp(-z)/\{1+\exp(-z)\} = \frac{\exp\left( \frac{-u_y   +  \beta_{j_0} X_{j_0}^{(L)}}{\sigma_{\varepsilon}} - \log 2 \right)}{1+ \exp\left( \frac{-u_y   +  \beta_{j_0} X_{j_0}^{(L)}}{\sigma_{\varepsilon}} - \log 2 \right)},
    \end{equation*}
    thus, similarly to the previous case,
    \begin{equation*}
    \text{logit}\{\text{Pr}(Y = 1 \mid X_{j_0}^{(L)} > u^{(L)})\} \approx -\log2 -\frac{u_Y}{\sigma_\varepsilon} + \frac{ \beta_{j_0}}{\sigma_\varepsilon}X_{j_0}^{(L)} = \tilde{\beta}_1+\tilde{\beta}_2X_{j_0}^{(L)}.
    \end{equation*}
\end{itemize}


\subsection*{H{\"u}sler--Reiss model}
In this framework we again obtain a probit model, since
\begin{align*}
    \text{Pr}\left(\tilde{Y}^{(L)} > u_Y^{(L)} \mid \sum_{j=1}^d \beta_j X_j^{(L)} > u^{(L)}\right)
    &= \text{Pr}\left(\tilde{\varepsilon}_{\bm \beta} > u_Y^{(L)}- \sum_{j=1}^d \beta_j X_j^{(L)}\right)\\ &= \Phi \left( \frac{\sum_{j=1}^d \beta_j X_j^{(L)} - u_Y^{(L)}+\lambda_{\bm \beta}/2}{\sqrt{\lambda_{\bm \beta}}}\right),
\end{align*}
meaning that $g^{-1}=\Phi$.

\subsection*{Regularly-varying linear framework}
For the Cauchy regression model the link function is the cauchit link. Indeed,
\begin{equation*}
\text{Pr}\left(\tilde{Y}^{(C)} >u_Y^{(C)}\mid \bm X^{(C)}\right) =\text{Pr}\left(\varepsilon > u_Y^{(C)} -  \sum_{j=1}^d \beta_jX_j^{(C)}\right)= F_{\text{Cauchy}} \left( \frac{\sum_{j=1}^d \beta_jX_j^{(C)} - u_Y^{(C)}}{\sigma_{\varepsilon}}\right),
\end{equation*}
and thus $g^{-1}=F_{\text{Cauchy}}$.

\section{Performance measures}\label{app:scores}
We recall here some details about some measures of predictive performance, which we use to compare different models. 
\subsection*{Proper scoring rules}
A scoring rule is called proper if the expected score, calculated with respect to a given distribution, is minimized when the forecast matches that distribution. It is strictly proper when this minimum is achieved only when the forecast distribution exactly coincides with the true distribution.
Proper scoring rules are commonly used in extreme value theory to assess the predictive performance of a model based on a forecast distribution $F$ and an observation $y$. Indeed, as discussed for instance in \cite{lerch2017forecaster}, common error-based metrics such as the MAE or MSE can be unreliable when comparing models based on extreme observations.
Two widely used scoring rules are the logarithmic score (LogS) \citep{good1952rational},
\begin{equation*}
\text{LogS}(F, y) = -\log f(y),
\end{equation*}
where $f$ denotes the density corresponding to $F$,
and the continuous ranked probability score (CRPS) \citep{matheson1976scoring},
\begin{equation*}
\text{CRPS}(F, y) = \int_{-\infty}^{\infty} \left\{ F(z) - \mathbbm{1}\{y \le z\} \right\}^2 \, \text{d}z.
\end{equation*}
In our regression setting, when the focus is on a binary response variable the logarithmic score is
\begin{equation*}
    \text{LogS}(\theta(\bm X^\star), y) = -[y\log\{\text{Pr}(Y = 1 \mid \bm X^\star)\} + (1-y)\log\{1-\text{Pr}(Y = 1 \mid \bm X^\star)\}].
\end{equation*}

\subsection*{AUC and AUPRC}
\vc{
Let TP, TN, FP, FN be the number of true positives, true negatives, false positives, and false negatives, respectively. 
The ROC curve is constructed by plotting the true positive rate, or sensitivity,
$$ \text{TPR}=\frac{ \text{TP}}{ \text{TP}+ \text{FN}}, $$
against the false positive rate,
$$ \text{FPR}=\frac{ \text{FP}}{ \text{FP}+ \text{TN}}, $$
for a set of possible classification thresholds. The AUC is then computed as the area under this curve. An AUC of 0.5 corresponds to discrimination no better than random ranking, whereas an AUC of 1 indicates perfect discrimination.}
\vc{
The precision–recall curve is instead constructed by plotting precision,
$$ \text{Precision}=\frac{ \text{TP}}{ \text{TP}+ \text{FP}}, $$
against recall,
$$ \text{Recall}=\frac{ \text{TP}}{ \text{TP}+ \text{FN}}, $$
across the same range of classification thresholds. The AUPRC summarizes the area under this curve, 
and it is particularly informative in settings with substantial class imbalance because it is directly sensitive to false-positive predictions and to the underlying event prevalence.
The expected AUPRC of a uninformative classifier is equal to the proportion of successes.}

\section*{Acknowledgments}
This work was initiated within the FIRE-RES project, funded by the European Union's Horizon 2020 research and innovation programme, and completed with support from the CLIMATHS project of the PEPR Maths-Vives project under the France 2030 investment plan coordinated by the national funding agency ANR. T. Opitz also acknowledges the support of the Chaire Geolearning funded by Andra, BNP-Paribas, CCR and SCOR Foundation.

{
\small
\bibliographystyle{abbrvnat}
\bibliography{biblio}
}

\end{document}